\documentclass[12pt]{article}

\usepackage[utf8]{inputenc}
\usepackage{setspace}
\usepackage{graphicx}
\usepackage{subcaption}
\usepackage{amsmath,amsfonts,amsthm, amssymb,ae, enumerate}
\usepackage{bm} 					
\usepackage{natbib}
\usepackage{algpseudocode,algorithmicx}

\usepackage{multirow}
\makeatletter
\renewcommand\paragraph{\@startsection{paragraph}{4}{\z@}%
            {-2.5ex\@plus -1ex \@minus -.25ex}%
            {1.25ex \@plus .25ex}%
            {\normalfont\normalsize\bfseries}}

\def\I{{\mathbb I}}

\newtheorem{prop}{Proposition}

\begin{document}

\title{\bf Multidimensional Bayesian IRT Model for Hierarchical Latent Structures}
\author{\bf{Juliane Venturelli$^{a}$ S. L., Flavio B. Gon\c{c}alves$^{a1}$, Dalton F. Andrade$^{b}$}}
\date{}

\maketitle

\begin{center}
$^a$ Universidade Federal de Minas Gerais, Brazil\\
$^b$ Universidade Federal de Santa Catarina, Brazil
\end{center}

\begin{abstract}
It is reasonable to consider, in many cases, that individuals' latent traits have a hierarchical structure such that more general traits are a suitable composition of more specific ones. Existing item response models that account for such hierarchical structure feature have considerable limitations in terms of modeling and/or inference. Motivated by those limitations and the importance of the theme, this paper aims at proposing an improved methodology in terms of both modeling and inference to deal with hierarchically structured latent traits in an item response theory context. From a modeling perspective, the proposed methodology allows for genuinely multidimensional items and all of the latent traits in the assumed hierarchical structure are on the same scale. Items are allowed to be dichotomous or of graded response. An efficient MCMC algorithm is carefully devised to sample from the joint posterior distribution of all the unknown quantities of the proposed model. In particular, all the latent trait parameters are jointly sampled from their full conditional distribution in a Gibbs sampling algorithm. The proposed methodology is applied to simulated data and a real dataset concerning the Enem exam in Brazil.
\end{abstract}

\section{Introduction}

Latent variables, also known as constructs, latent traits or factors, are features that can not be directly measured, but that can be indirectly measured by some manifests, such as responses to items. Intelligence, personality traits and depression are some examples of construct. Typically, the main aim of psychometric tests is to create a scale to measure latent traits of individuals. It is commonly the case that more than one specific trait is measured by a test and those can be seen as different spectra of a common general trait.
Take for instance a language construct which itself can be perceived as a higher dimension with subdimensions, such as reading and writing abilities. The reading process in turn could also be seen as a higher dimension formed by subdomains such as interpretation, evaluation and retrieval of information. It may be the case that the researcher is interested not only in measuring the general dimension, say language ability, but also in measuring those other subdomains in order to have a more detailed characterization of the subjects participating in a study. Going on the other direction, from more specific to more general latent traits, consider university admission tests like SAT in the USA and Enem in Brazil, in which candidates are submitted to tests in specific areas, like maths, languages and sciences. In addition to measuring the specific traits, it also useful, for example in order to better classify candidates, to measure one general trait that is in a way a suitable combination of the specific ones.

In academic tests like the two cited above, it is expected to have a significant correlation among the different specific tests. Indeed, exploratory analyzes of the results from Enem, in which the specific traits are obtained independently for each test, show a significant correlation among the performances of the individuals in the specific tests (see Table \ref{tab:MatrixCorr1} below). It is then important to consider models that suitably account for those features. Naturally, that demands complex modeling structures and, consequently, carefully designed methodologies to perform efficient statistical inference. A suitable hierarchical modeling of the latent traits ought to result in more refined/precise measurements than considering one common general trait to directly explain the performance in all the specific tests. Moreover, given the natural correlation among the specific areas, further improvement is achieved if items are allowed to be multidimensional, in the sense of measuring more than one specific ability. For instance, an item in the maths test may have a long and relatively complex text such that text interpretation skills influence on the response given to that item. Also, a physics item may require considerably good math skills to be solved.

The aim of this paper is to propose an statistical methodology consisting of: 1. an item response theory (IRT) model that is able to efficiently accommodate a flexible hierarchical structure for the latent traits measured by multidimensional dichotomous and graded response items, that considers a 3-parameter formulation to model the dichotomous items; 2. an MCMC algorithm to perform Bayesian inference for the proposed model that targets the posterior distribution of all the unknown quantities in the model and aims at being as computationally efficient as possible. One distinguish property of the proposed algorithm is the joint sampling of all the latent traits of all the individuals. Also, another important feature of the proposed model is the fact that all the latent traits (from all the hierarchical levels) are in the same scale and model identifiability, which is a complex issue in IRT models, is achieved without imposing unreasonable restrictions to the model parameters.

IRT models with hierarchical structures for the latent traits have already been considered in the literature. \cite{de2004higher} proposed a higher order latent trait model for cognitive diagnosis considering only dichotomous latent traits. \cite{sheng2008bayesian} proposed a Bayesian approach to estimate both the general and specific abilities using a multiunidimensional IRT model on the item level. However, their parametrization does not fix the scale of the latent traits whilst having distinguish scales for each trait. This feature seriously jeopardizes parameter/model interpretation. \cite{Torre2009} also proposed a model in which the overall and specific latent traits are simultaneously estimated. Nevertheless, item parameters are assumed to be known, which is typically an unrealistic assumption. \cite{huang2013higher} also proposed a higher order model extending the work of \cite{Torre2009} for more than two levels and also simultaneously estimating item parameters, but under some inference drawbacks. More recently, \citet{delatorre2019} proposed a higher order IRT model to estimate traits of participants from multiple studies that allows for more than one higher order trait. Conditional on the first-level traits, the responses are modeled using a multiunidimensional 2PL model.

None of the references above have addressed the inference process in detail, despite its high complexity. This regards issues related to model identifiability, parameter interpretation and convergence of the algorithms. Another considerable limitation of the existing methodologies is the fact that they consider, at most, a multiunidimensional structure at the item level, meaning that items measure only one specific latent trait on the first level. Naturally, considering genuinely multidimensional items ought to significantly increase the complexity of the inference process. To the best of our knowledge, this is the first work in the literature to efficiently deal with multidimensional IRT models with complex hierarchical structures for the latent traits, considering both dichotomous and graded response items with a 3-parameter model for the dichotomous items.

Important issues regarding modeling and inference are explored through simulated examples and a real dataset from the Enem exam is analized to illustrate the applicability of the proposed methodology.

This paper is organized as follows. Section 2 presents the proposed hierarchical MIRT model and Section 3 presents the MCMC algorithm. Results from simulated and real examples are presented in Sections 4 and 5, respectively.

\section{Model specification}

Let $\bm{\theta}^{(k)}_{j}=(\theta^{(k)}_{1j},\ldots,\theta^{(k)}_{Q_kj})'$ denote vector of latent traits of subject $j, j = 1, \dots, J$, on hierarchical level $k$, $k = 1, \dots, K$, $q = 1, \dots, Q_k$, where $Q_k$ denotes the number of latent traits on level $k$. Now define $\bm{\theta}_{j} =(\bm{\theta}^{(1)}_{j},\dots, \bm{\theta}^{(K)}_{j})$, $\bm{\theta} = (\bm{\theta}_1 \ldots \bm{\theta}_J)$ and the analogous notation replacing $\theta$ by $\epsilon$. The following hierarchical structure is considered for the latent traits:
\begin{equation}\label{eq:StructHier}
\bm{\theta}^{(k)}_{j} = \bm{\lambda}^{(k+1)}\bm{\theta}^{(k+1)}_{j} + \bm{\epsilon}^{(k)}_{j},\qquad k=1,\dots,K-1,
\end{equation}
where $\bm{\lambda}^{(k+1)}$ is the $Q_{k} \times Q_{k+1}$ coefficient matrix with entries $\lambda{qq'}^{(k+1)}$ that relate $\bm{\theta}^{(k)}_{j}$ to $\bm{\theta}_{j}^{(k+1)}$, $\forall$ $j$, and the $\epsilon^{(k)}_{qj}$'s are random errors. We impose a constraint to matrices $\bm{\lambda}^{(k+1)}$ so that each $\theta^{(k)}_{qj}$ is explained by only one trait from level $(k+1)$. This implies that each row of the matrix $\bm{\lambda}^{(k+1)}$ has only one non-null element. Also, in order to identify the model, we impose the standard factor analysis restriction that each trait on an upper level explains at least three traits on its immediate lower level.

In the hierarchical structure defined in (\ref{eq:StructHier}), it is the lowest level traits $\bm{\theta}^{(1)}$ that will be directly measured by the responses given to the items. The factor analysis modeling structure combines those traits to define the upper level traits such that explains the most the lower level ones, given the specifications of the researcher. This way, a latent trait on an upper level ought to be interpreted as the trait the best characterizes the general ability defined as a composition of the respective lower level abilities.

Under a Bayesian approach, the full prior specification of the latent traits is completely specified by also assuming the following:
\begin{description}
	\item[A1]
	\begin{equation}\label{eq:ThetaK}
	\theta^{(K)}_{qj} \sim  N(0, 1), \quad \forall q,\;\forall j.
	\end{equation}
	
	\item[A2]
	\begin{equation}\label{eq:ErroK}
	(\epsilon^{(k)}_{qj}|\lambda_{q\cdot}^{(k+1)}) \sim N(0, 1 - \sum^{Q_{k+1}}_{q'= 1}(\lambda_{qq'}^{(k+1)})^2).
	\end{equation}
	
	\item[A3]
	\begin{equation}
	\pi(\bm{\theta}, \bm{\lambda)} =  \biggl[\prod_{j=1}^J \prod_{q=1}^{Q_K} \pi(\theta^{(K)}_{qj})\biggr] \biggl[ \prod^J_j\prod_{k=1}^{K-1} \prod_q^{Q_k}  \pi(\theta_{qj}^{(k)}|\lambda_{qq'}^{(k+1)}, \theta_{q'j}^{(k+1)})\biggr]\biggl[\prod_{k=2}^K\prod_{q=1}^{Q_{k}} \prod_{q'=1}^{Q_{k+1}}  \pi(\lambda_{qq'}^{(k)})\biggr].
	\end{equation}
\end{description}
We adopt Uniform(-1,1) priors for all the coefficients $\lambda_{qq'}^{(k)}$. Note that, given the restrictions imposed to matrices $\bm{\lambda}^{(k+1)}$, only one parcel in the sum $\sum^{Q_{k+1}}_{q'= 1}(\lambda_{qq'}^{(k+1)})^2$ is different from zero.

The prior distribution defined by equation (\ref{eq:StructHier}) and the specifications in \textbf{A1}-\textbf{A3} has two main theoretical advantages. First, it will establish the identification of the full IRT model to be presented further ahead by setting the scale of all the latent traits and, second, it sets those scales to be the same, as stated in the following proposition.
\begin{prop}\label{prop1}
	The prior distribution defined in (\ref{eq:StructHier}) and \textbf{A1}-\textbf{A3} implies that the marginal prior distribution of each $\theta^{(k)}_{qj}$, for all $k$, $q$ and $j$,  is normal with mean zero and variance one.
\end{prop}
The proof of Proposition \ref{prop1} is presented in Appendix A.

Now note that the prior on $\bm{\theta}$ also implies that, conditional on the level $(k+1)$ latent traits and $\lambda^{(k+1)}$, the $Q_k$ traits of subject $j$ on level $k$ are independent with $(\theta_{qj}^{(k)}|\bm{\lambda}^{(k+1)},\bm{\theta}_{j}^{(k+1)})\sim N(\bm{\lambda}_{q\cdot}^{(k+1)}\bm{\theta}_{j}^{(k+1)},1-\sum^{Q_{k+1}}_{q'= 1}(\lambda_{qq'}^{(k+1)})^2)$. The conditional independence among the latent traits of different subjects is also implied.

We embed the hierarchical prior structure on the latent traits defined above into an IRT model with multidimensional dichotomous and graded response items. .

Assume that a random sampling of $J$ subjects is submitted to a test composed by $I$ items. Denote by $\bm{Y} = \{Y_{ij}\}_{I \times J}$, the $I \times J$ matrix of all responses to the test, where $Y_{ij}$ is the indicator of a correct response given to a dichotomous item $i$ or the score of the response to a graded item $i$ by individual $j$, for $i = 1,\ldots, I$ and $j = 1, \ldots,J$. The model for the dichotomous items is defined as follows.
\begin{equation}\label{eq:ICC}
\begin{split}
Y_{ij} \rvert p_{ij} &\overset{ind}{\sim} Bernoulli(p_{ij}),\\
p_{ij} &= c_{ij} + (1-c_{ij})\Phi( \textbf{a}_i\bm{\theta}^{(1)}_j - b_i),\\
\end{split}
\end{equation}
where $\textbf{a}_i = (a_{i1} \dots a_{iQ_{1}})$ is the vector of discrimination parameters for item $i$ and $b_{i}$ and $c_i$ are the difficulty and pseudo guessing parameters, respectively, for item $i$. The link function $\Phi(.)$ is the c.d.f. of the standard Normal distribution. Differently from an unidimensional IRT model, the proposed model allows item $i$ to measure more than one latent trait. In order to identify the model in (\ref{eq:ICC}) some additional restrictions must be imposed. One reasonable possibility is to make $a_{iq} = 0$, for $i = 1,\dots, Q_1-1$, and $q = i + 1,\dots, Q_1$ \citep[see][]{beguin2001mcmc}.

The Graded Response Model (GRM) of \citet{samejima1969estimation} is used to model the graded response items in which the possible responses are classified into ordered categories. Assuming that item $i$ has $M_i$ categories, $M_i-1$ location parameters $b_{im}$ are considered such that
\begin{equation}\label{eq:ordemb}
-\infty = b_{i0} < b_{i1} <  b_{i2} < \dots <  b_{i,M_i-1} < b_{i,M_i} = \infty.
\end{equation}
Defining $\bm{b}^G_i = (b_{i1}^G, \dots, b_{i,M_i-1}^G)$, $\bm{b^G} = (\bm{b}^G_1,\dots, \bm{b}^G_I)$, $\bm{a}^G_i = (a^G_{i1}, \dots, a^G_{i,Q_1})$ and $\textbf{a}^G = (\bm{a}^G_1,\dots, \bm{a}^G_I)$, the GRM assumes that the probability that a subject $j$ receives a score $m$ in item $i$ is
\begin{equation}\label{eq:samejima}
P(Y_{ij} = m) = \Phi(b_{i,m} - \bm{a}^G_{i}\bm{\theta}_{j} ) - \Phi(b_{i,m-1} - \bm{a}^G_{i}\bm{\theta}_{j}).
\end{equation}

The likelihood function of the full model factorizes into the product of the likelihood for the dichotomous items and likelihood for the graded response items. Given the conditional independence among the responses implied in (\ref{eq:ICC}), the former is given by
\begin{equation} \label{LilPM}
L(\bm{\theta}^{(1)}, \textbf{a}, \textbf{b}|\textbf{Y}) = \prod^J_{j=1}\prod^I_{i=1}p_{ij}^{y_{ij}}(1 - p_{ij})^{1 - y_{ij}},
\end{equation}
where $\textbf{a} = (\textbf{a}_1 \dots \textbf{a}_I)$ and $\textbf{b} = (b_1, \dots, b_I)$. Conditional independence among the responses of the graded response items is also assumed, leading to the following likelihood
\begin{eqnarray} \label{LikeGraded}
L_2(\bm{\theta}^{(1)}, \textbf{a}^G, \textbf{b}^G|\textbf{Y}) = \prod^J_{j=1}\prod^I_{i=1}\prod^{M_i}_{m=1} [\Phi(b_{im} - \bm{a}^G_{i}\bm{\theta}^{(1)}_{j} ) - \Phi(b_{im-1} - \bm{a}^G_{i}\bm{\theta}^{(1)}_{j})]^{\I(Y_{ij}=m)},
\end{eqnarray}
where $\I(Y_{ij}=m) = 1$ is the indicator of subject $j$ receiving a score $m$ in item i.

We assume independent normal priors for all the item parameters considering the restriction in (\ref{eq:ordemb}). We set the notation $(\textbf{a}_i, b_i) \sim N(\bm{\mu}, \bm{\Lambda})$, $(\textbf{a}_{i}^{G})\sim N(\bm{\mu}_a, \bm{\Lambda}_a)$ and adopt $b_{im}\sim N(0,1)$.

\section{Model Estimation}\label{secmcmc}

Bayesian estimation for the model presented in the previous section is carried out via MCMC, more specifically, a Gibbs sampling algorithm to sample from the joint posterior distribution of all the unknown quantities in the model. In order to facilitate the calculations and to make direct simulation from as many full conditional distributions as possible in the Gibbs sampling feasible, we introduce three sets of auxiliary variables. 

For all the dichotomous items and all the subjects responding these, we consider the following auxiliary variable proposed by \cite{bambirra2018bayesian}.
\begin{equation}
 Z_{ij} \sim Bernoulli(c_i), \qquad(X_{ij}|Z_{ij}) \sim N(\textbf{a}_i\bm{\theta}_j - b_i, 1)\I(Z_{ij}=0) + \delta_0\I(Z_{ij}=1),
\end{equation}
where $\I$ is the indicator function and $\delta_0$ is a point-mass at 0, that is, $P(X_{ij} = 0 | Z_{ij} = 1) = 1$. We then make
\begin{equation}\label{FlavioBarbara}
Y_{ij}=\left\{
\begin{array}{ll}
1,&\mbox{if}\quad (Z_{ij}=1) \quad \mbox{or} \quad (Z_{ij}=0, X_{ij} \geq 0),\\
0,&\mbox{if}\quad (Z_{ij}=0, X_{ij} < 0),
\end{array}
\right.
\end{equation}
where $\textbf{Z} = \{Z_{ij}\}$, $\textbf{X} = \{X_{ij}\}$, $\forall$ $i,j.$

For the graded response items, we consider the auxiliary variables proposed in \citet{johnson2006ordinal} so that
\begin{equation}\label{hierqPolitomico}
    Y_{ij}=\left\{
    \begin{array}{ll}
    0,&\mbox{if}\quad  -\infty < X_{ij}^G < b_{i1},\\
    1,&\mbox{if}\quad b_{i1} < X_{ij}^G < b_{i2}, \\
    \vdots\\
    M_i, &\mbox{if}\quad b_{i,M_i-1} < X_{ij}^G < \infty,
    \end{array}
    \right.
\end{equation}
where $X_{ij}^G\overset{ind}{\sim} N(\textbf{a}^G_i\bm{\theta}_j^{(1)},1)$.

The augmented model can be shown to be equivalent to the original model upon marginalization w.r.t. the auxiliary variables. Defining $\Psi$ as the vector of all the unknown quantities in the augmented model, all the calculations to devise the MCMC algorithm to sample from the posterior of $(\Psi|Y)$ are based on the joint density of $Y$ and $\Psi$ which is as follows.
\begin{equation}\label{Posterior}
\begin{split}
\pi(\textbf{Y},\bm{\Psi}) =&
\pi(\textbf{Y}|\textbf{X},\textbf{Z})\pi(\textbf{X}|\textbf{Z},\textbf{a},\textbf{b},\bm{\theta}^{(1)})\pi(\textbf{X}^G|\textbf{a}^G,\textbf{b}^G,\bm{\theta}^{(1)})\pi(\textbf{Z}|\textbf{c}) \\
&\times  \prod_{k=2}^{K}\pi(\bm{\theta}^{(k-1)}|\bm{\theta}^{(k)}, \bm{\lambda}^{(k)})\pi(\textbf{a},\textbf{b})\pi(\bm{\lambda})\pi(\bm{\theta}^{(K)})\pi(\textbf{c})\pi(\textbf{b}^G)\pi(\textbf{a}^G).
\end{split}
\end{equation}

The blocking scheme of the Gibbs sampling algorithm is
$$(\textbf{X},\textbf{Z}) \quad (\textbf{a},\textbf{b}) \quad  ({\bm{\lambda}}_{q\cdot}^k) \quad (\bm{\theta}) \quad (\bm{c}) \quad (\bm{X}^G) \quad (\bm{a}^G) \quad (\bm{b}_{i}^G),$$
where, for each pair $(q,k)$, we have a block $\bm{\lambda}_{q\cdot}^{(k)}$ and, for each $i$, we have a block $\bm{b}_{i}^G$. The high dimensionality of most of the blocks in te sampling scheme above leads to a computationally efficient algorithm. We highlight the joint sampling of $\bm{\theta}$ which includes the traits in all levels and for all individuals. Each block $\bm{\lambda}_{q}^{(k)}$ is actually composed by its non-null coordinate and these parameters cannot be sampled directly from their respective full conditional distributions. Therefore, they are sampled in Metropolis Hastings steps. Another important aspect of the MCMC algorithm is that parameters $\bm{b}^G$ are typically highly correlated to the auxiliary variable $\bm{X}^G$. In order to solve this problem, we propose a reparametrization of parameters $\bm{b}^G$ and an algorithm to sample each vector $\bm{b}_{i}^G$ using two different Metropolis Hastings steps. Details about this and all the other steps of the MCMC algorithm are presented in Appendix B.

\section{Simulated examples}

We present three simulated examples to investigate the efficiency of the methodology proposed in this paper. We analyze the efficiency of the proposed methodology to estimate the parameters under different specifications for the number of subjects. The data is simulated from a model with two levels of latent traits - four traits on the first level and one on the second level. The true coefficient values are $\bm{\lambda}$ = ($\lambda_1$ = 0.95, $\lambda_2$ = 0.90, $\lambda_3$ = 0.85,$\lambda_4$ = 0.80). Each trait is measured by 45 unidimensional dichotomous items and all the pseudo-guessing parameters are set to be zero. A model with the same properties is used to fit the data.

Table \ref{tab:2kStud} presents the posterior statistics for $\bm{\lambda}$ for sample sizes consisting of 500, 2k and 5k subjects. Notice that the $\bm{\lambda}$ parameters are satisfactorily recovered for all sample sizes with posterior precision increasing with the sample size. Moreover, the higher is the true $\lambda$ value, the smaller is its posterior standard deviation. Standard diagnostics strongly suggest the convergence of the MCMC algorithm. Figure \ref{graf:Simulacao7_500_thetas} (see Appendix C) shows the posterior mean versus true values for the first and second level $\theta$, and Figure \ref{graf:Simulacao7_500_ParItens} (see Appendix C) shows the posterior mean versus true values for the item parameters. Notice that even for a small sample size as 500 all model parameters are fairly well recovered.

\begin{table}[h!]
	\centering
	\begin{tabular}{ccccccc}\hline
		Simulation&Sample Size &  $\lambda$   & Mean & StandDev & $CI_{0.025}$& $CI_{0.975}$  \\\hline
		&\multirow{3}{*}{500}&0.95& 0.957& 0.010& 0.937& 0.974\\
		1		 &&0.90& 0.905& 0.013& 0.879& 0.928 \\
		& &0.85& 0.852& 0.017& 0.817& 0.883 \\
		& &0.80& 0.836& 0.018& 0.797& 0.869 \\\hline
		&\multirow{3}{*}{2.000} &                0.95& 0.946& 0.005& 0.937& 0.956 \\
		2		& &0.90& 0.904& 0.006& 0.892& 0.917 \\
		& &0.85& 0.865& 0.008& 0.849& 0.879 \\
		& &0.80& 0.791& 0.010& 0.770& 0.810 \\\hline
		&\multirow{3}{*}{5.000} &0.95& 0.949& 0.003& 0.942& 0.955 \\
		3            &         & 0.90 &0.900& 0.004& 0.891& 0.908 \\
		&        &0.85& 0.853& 0.005& 0.842& 0.862   \\
		&       &0.80& 0.794 & 0.006& 0.781& 0.807 \\ \hline
	\end{tabular}
	\caption{Posterior statistics for $\bm{\lambda}$ for different subject sample sizes.}
	\label{tab:2kStud}
\end{table}

We now study the impact of not considering a hierarchical structure when there is one and vice versa. In simulation 4, data was generated from a hierarchical model, such that $\bm{\lambda} = (\lambda_1 = 0.95, \lambda_1 = 0.90, \lambda_1 = 0.85, \lambda_1 = 0.80 )$, but the estimated model ignored the hierarchical structure imposing that $\lambda = (0, 0, 0, 0)$, that is, four unidimensional unrelated tests. In simulation 5, the dataset was generated considering uncorrelated subtests, that is, $\lambda = (0, 0, 0, 0)$, but the estimated model allowed a hierarchical structure. In simulation 6, dataset was generated from an unidimensional model, that is, all $\lambda = 1$, but a hierarchical structure was allowed to fit the data, assuming four traits on the first level and one on the second level. In simulation 7, the dataset was generated with a hierarchical structure, such that $\bm{\lambda} = (\lambda_1 = 0.95, \lambda_1 = 0.90, \lambda_1 = 0.85, \lambda_1 = 0.80 )$, and the same structure was considered to estimate parameters. For simulations 4 to 7, pseudo-guessing parameters were set to be zero and only dichotomous items (45 for each first level trait) were considered, with 5k subjects. Also, in these four studies, item parameters were fixed at their real value to isolate noise effects.

Notice that, when all $\lambda$ = 0 the whole test is composed by uncorrelated subtests and when all $\lambda$ = 1 the test is unidimensional. Therefore, both models are particular cases of the proposed hierarchical model. Moreover, one can use the proposed model to test unidimensionality or uncorrelated latent traits, as we show in these studies. This means that the proposed model can be used even when the researcher is not sure about the existence of a hierarchical structure.

\begin{figure}[h!]
	\centering
	\begin{subfigure}[b]{0.4\textwidth}
		\includegraphics[width=\textwidth]{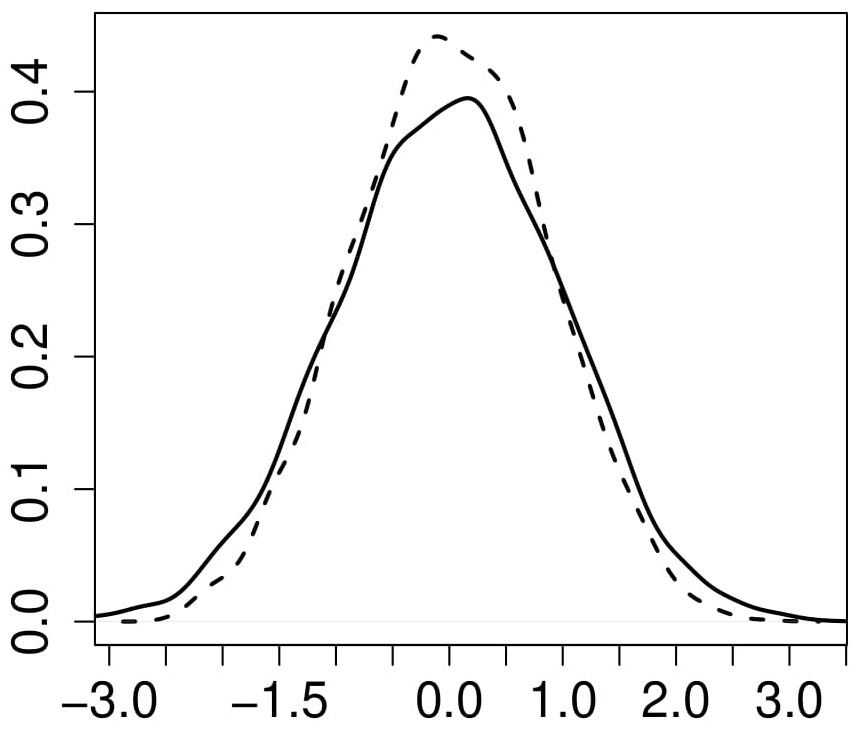}
	\end{subfigure}
	~
	\begin{subfigure}[b]{0.4\textwidth}
		\includegraphics[width=\textwidth]{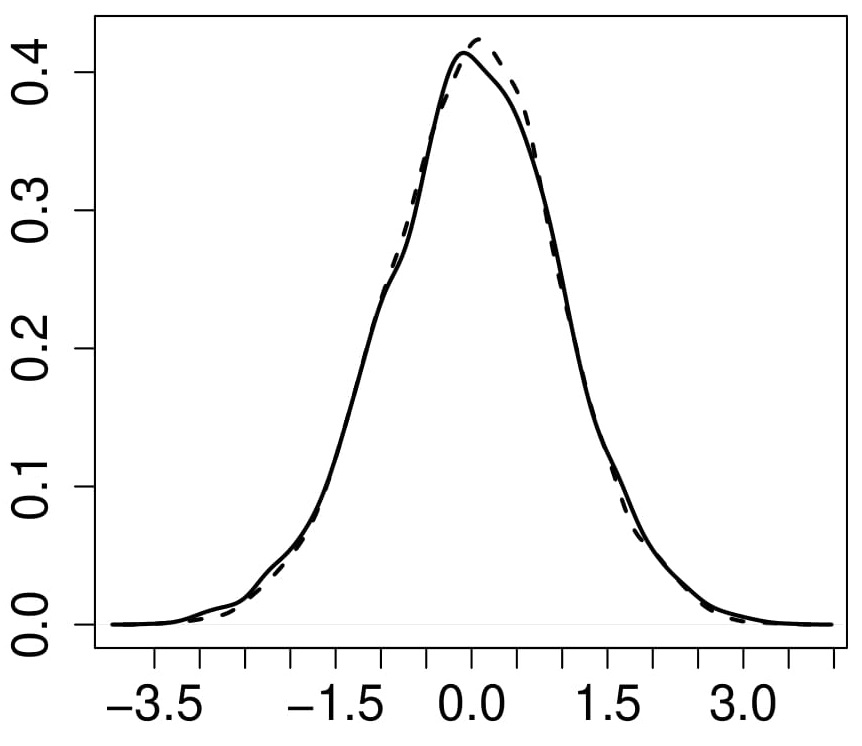}
		\end{subfigure}
	\caption{Density plot of real (dotted line) and estimated (solid) values for $\theta^{(2)}$: simulations 4 (left) and 7 (right).}
	\label{tab:Densidades}
\end{figure}

Table \ref{tab:RMSE} shows the Root Mean Squared Error (RMSE), $\sqrt{\sum_{j=1}^J(\hat \theta^{(k)}_{qj}-\theta^{(k)}_{qj} )^2/J}$ for the latent traits. The lowest RMSE is for simulation 6, where dataset was unidimensional, i.e., $\lambda = 1$ for all subtests. This is expected as all the 180 items are basically measuring only one latent trait. The RMSE for simulations 4 and 5 were slightly higher than the RMSE for simulation 7. Also, RMSE for second level $\theta_1^{(2)}$  in simulation 4 was obtained by averaging the posterior mean of the four first level latent traits. We highlight the fact that the $\theta^{(2)}$ parameters are better estimated under the hierarchical approach. This is also verify in Figure \ref{tab:Densidades} which presents the density of real and estimated values (posterior mean) of $\theta_1^{(2)}$ for both simulations 4 and 7. As expected, the model that considers the hierarchical structure recovers $\theta^{(2)}$ better than a simple average. This is expected mostly when the values of $\bm{\lambda}$ greatly differ, meaning that lower level traits are not equally related to the higher latent traits. When the subtests form an unidimensional test, indicating almost perfect correlation among first level traits, all $\lambda$'s are estimated near one. When there are uncorrelated subtests, all $\lambda$ are estimated near zero (see Table \ref{tab:lambda1}).
\begin{table}[h!]
	\centering
	\begin{tabular}{cccccc}
		\hline
		Simulation & $\theta_1^{(1)}$ & $\theta_2^{(1)}$ & $\theta_3^{(1)}$ & $\theta_4^{(1)}$ &$\theta_1^{(2)}$  \\ \hline
		4   &  0.255& 0.238& 0.251& 0.248& 0.313  \\
		5   &  0.250& 0.248& 0.240& 0.249& - \\
		6   &  0.127& 0.126& 0.126& 0.127& 0.1260\\
		7   &  0.221& 0.225& 0.221& 0.241& 0.271                 \\ \hline
	\end{tabular}
	\caption{Root mean squared error.}
	\label{tab:RMSE}
\end{table}

Although the model's structure is considerably complex, the studies presented above show that item and subject's parameters are satisfactorily recovered. As it was noticed before, other models, including the unidimensional one, are nested in the proposed hierarchical model. We highlight the fact that despite the high complexity of the model, its identifiability is achieved whist preserving its flexibility and practical interpretation.

\begin{table}[h!]
	\centering
	\begin{tabular}{ccccccc}
		\hline
		Simulation &Real&  Mean&    StandDev&     $CI_{0.025}$& $CI_{0.975}$  &  Range\\ \hline
		5 & 0 &-0.023 &0.123 &-0.280& 0.212& 0.492\\
		& 0&  0.023& 0.146& -0.248 &0.289& 0.538\\
		& 0& -0.022 &0.120& -0.338& 0.168& 0.506\\
		&0 &-0.001 &0.121& -0.255 &0.212&0.466\\\hline
		6&1.00& 0.9992& 2e-04& 0.9986& 0.9995& 0.0010\\
		&1.00& 0.9999& 0e+00& 0.9998& 0.9999& 0.0001\\
		&1.00& 0.9992& 5e-04& 0.9982& 0.9997& 0.0015\\
		&1.00& 0.9985& 4e-04& 0.9976& 0.9991& 0.0016\\\hline
	\end{tabular}
	\caption{Posterior statistics for $\bm{\lambda}$ - simulations 5 and 6.}
	\label{tab:lambda1}
\end{table}

We also present an example - simulation 8, for an exam with both dichotomous and graded items - some being multidimensional, and also considering the pseudo-guessing parameter for all items. The hierarchical structure is composed by five first level traits and one second level trait such that $\bm{\lambda}$ = ($\lambda_1$ = 0.95, $\lambda_2$ = 0.90, $\lambda_3$ = 0.85, $\lambda_4$ = 0.80, $\lambda_5$ = 0.80). The dataset consists of 10k subjects, 180 dichotomous items and five graded items. We first relate 45 items to each trait and then: five items of trait $\theta_2^{(1)}$ are related to $\theta_1^{(1)}$, five items of trait $\theta_3^{(1)}$ are related to $\theta_1^{(1)}$, six items of trait $\theta_4^{(1)}$ are related to $\theta_1^{(1)}$, and one item of $\theta_5^{(1)}$ (graded item) is related to $\theta_1^{(1)}$. This structure is considered in order to replicate the structure in the real data analysis to be presented in Section \ref{realanalysis}. The pseudo-guessing of all items were randomly selected from a Uniform(0,0.2) distribution.

\begin{figure}[h!]
	\centering
	\begin{subfigure}[b]{0.33\textwidth}
		\includegraphics[width=\textwidth]{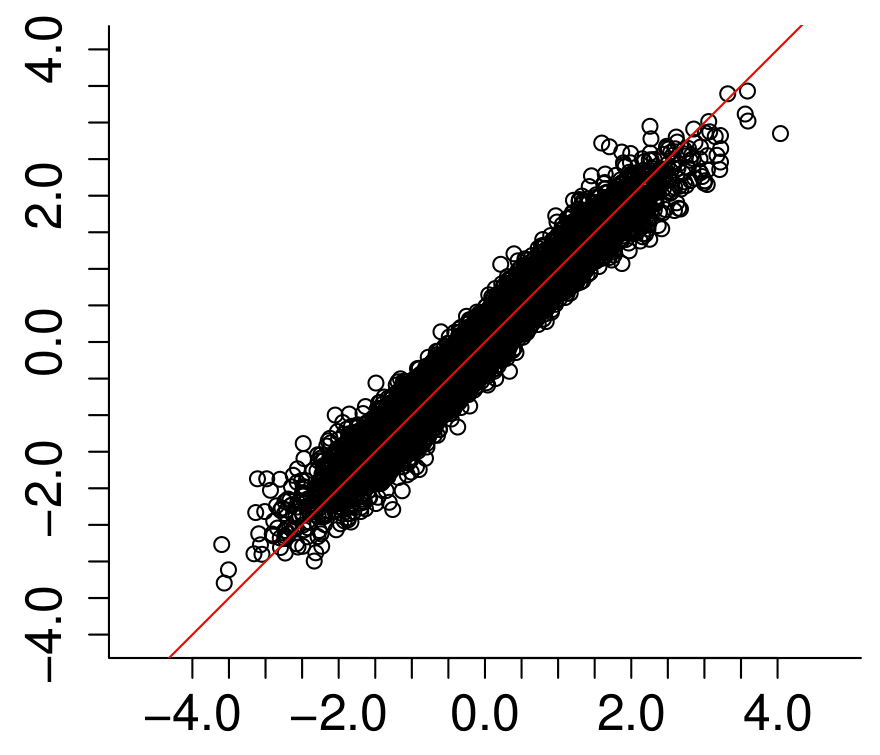}
		\caption{$\theta^{(1)}_1$}
	\end{subfigure}
	~
	\begin{subfigure}[b]{0.33\textwidth}
		\includegraphics[width=\textwidth]{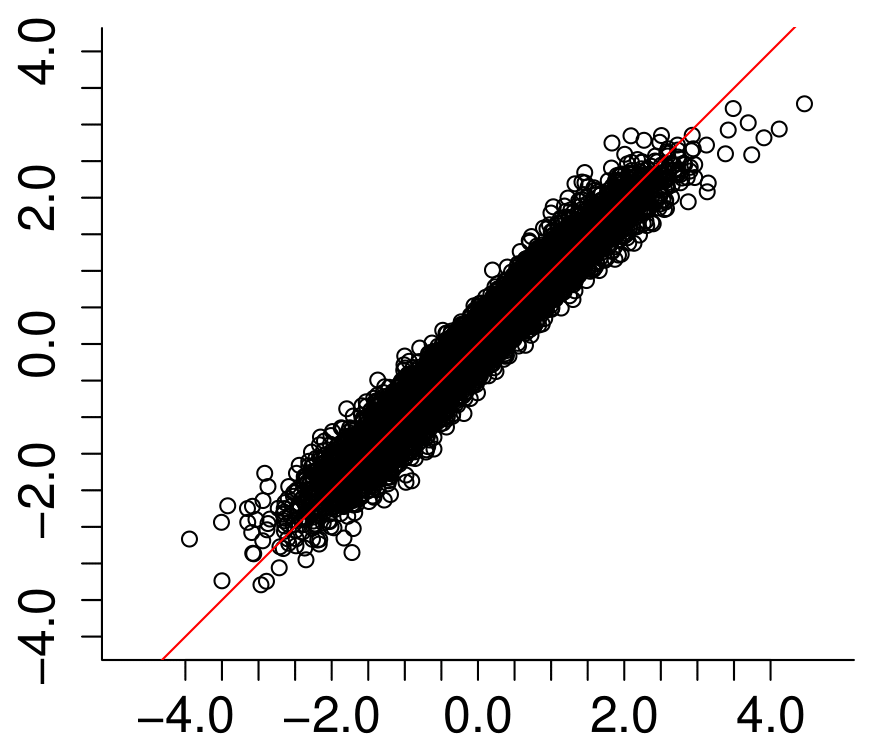}
		\caption{$\theta^{(1)}_2$}
	\end{subfigure}
	~ 
	\begin{subfigure}[b]{0.33\textwidth}
		\includegraphics[width=\textwidth]{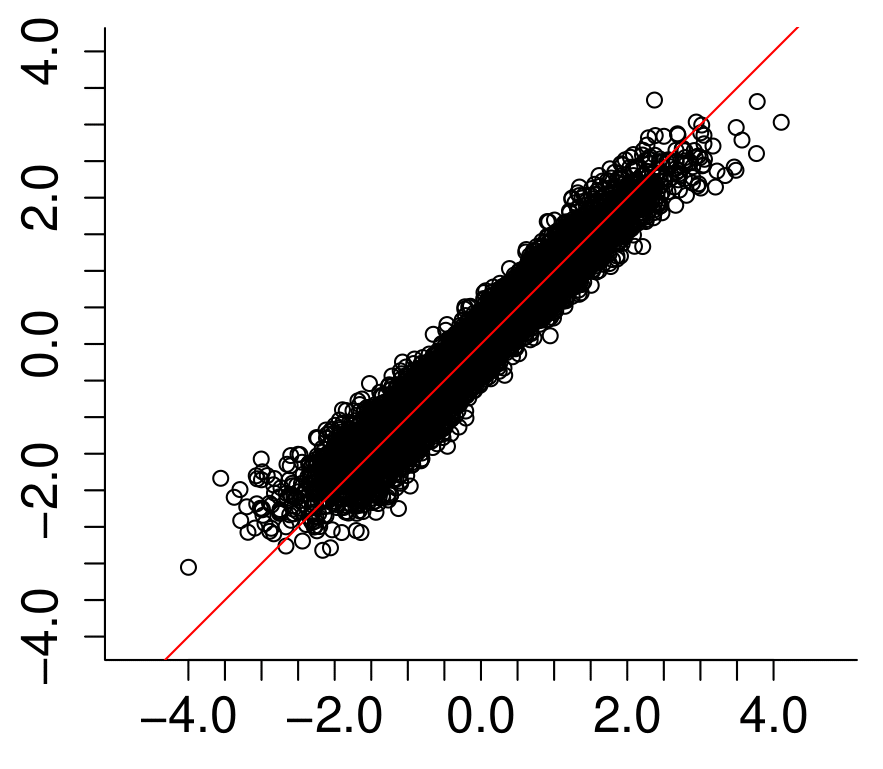}
		\caption{$\theta^{(1)}_3$}
	\end{subfigure}
	~
	\begin{subfigure}[b]{0.33\textwidth}
		\includegraphics[width=\textwidth]{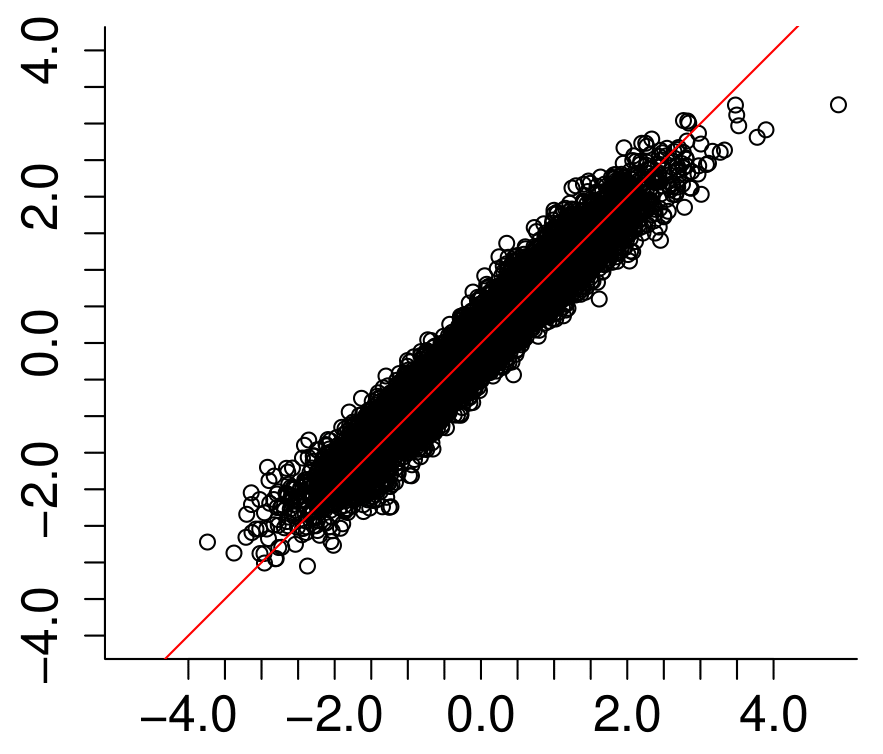}
		\caption{$\theta^{(1)}_4$}
	\end{subfigure}
	~ 
	\begin{subfigure}[b]{0.33\textwidth}
		\includegraphics[width=\textwidth]{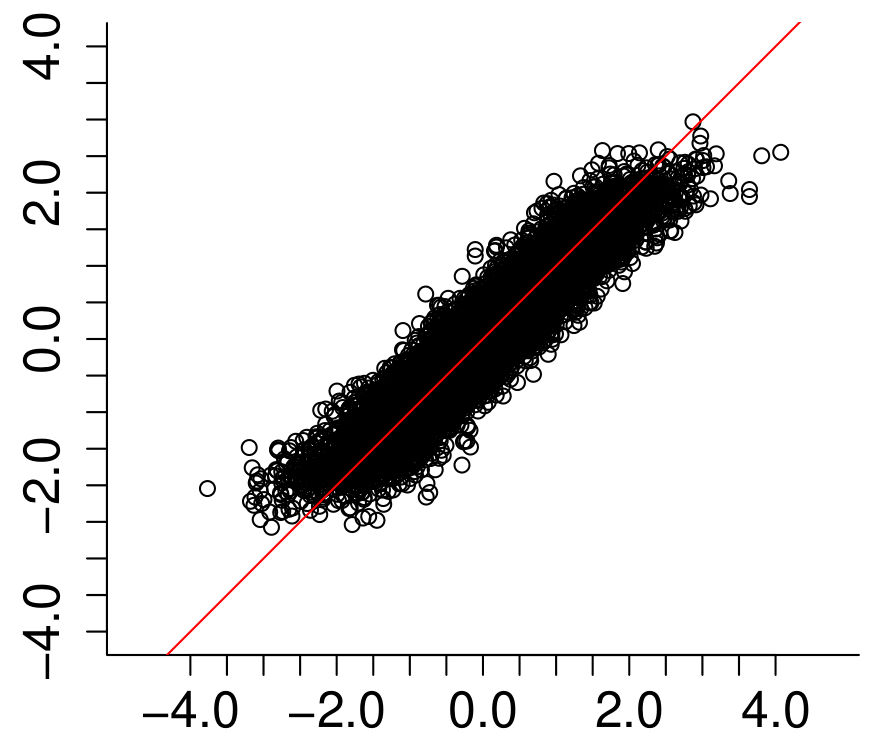}
		\caption{$\theta^{(1)}_5$}
	\end{subfigure}
	~
	\begin{subfigure}[b]{0.33\textwidth}
		\includegraphics[width=\textwidth]{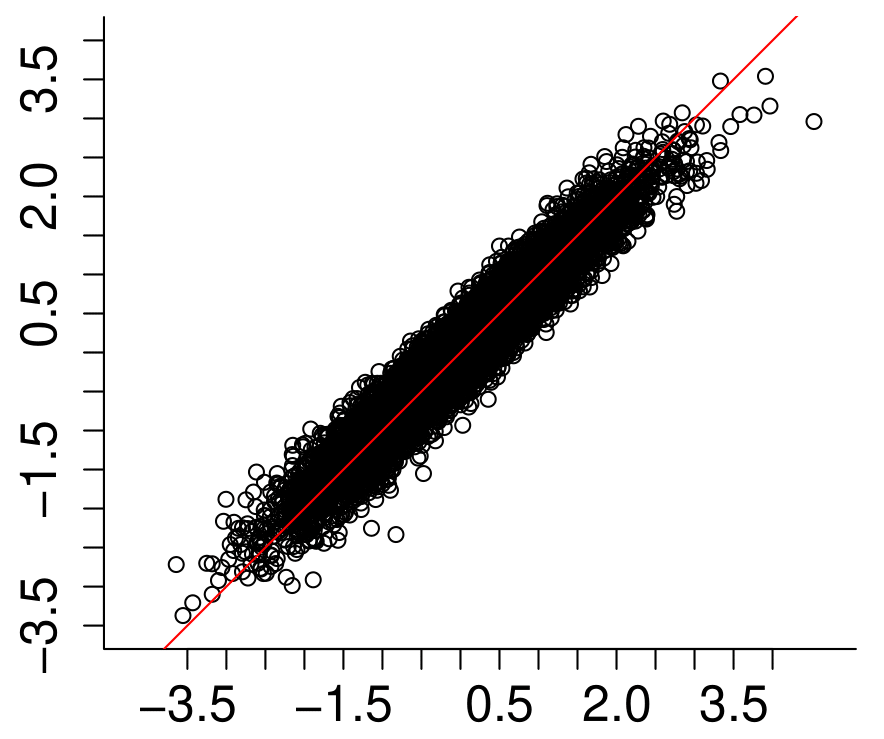}
		\caption{$\theta^{(2)}_1$}
	\end{subfigure}
	\caption{True value (x-axis) versus posterior mean (y-axis) for latent traits (simulation 8).}
	\label{graf:Simulation9_10k}
\end{figure}

\begin{figure}[h!]
	\centering
	\begin{subfigure}[b]{0.25\textwidth}
		\includegraphics[width=\textwidth]{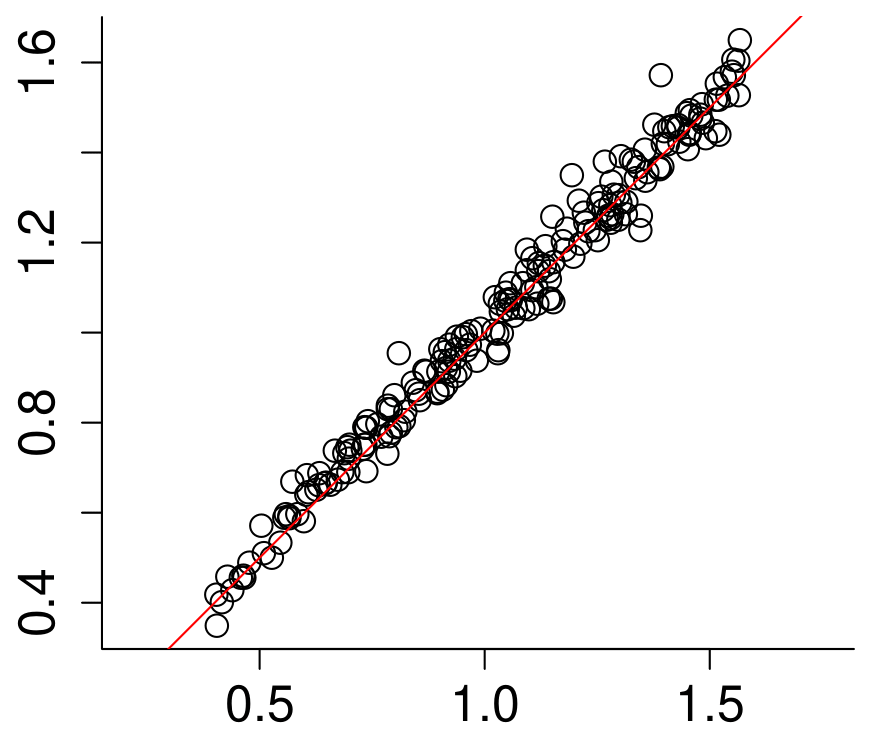}
		\caption{$a_i$ parameters.}
	\end{subfigure}
	~
	\begin{subfigure}[b]{0.25\textwidth}
		\includegraphics[width=\textwidth]{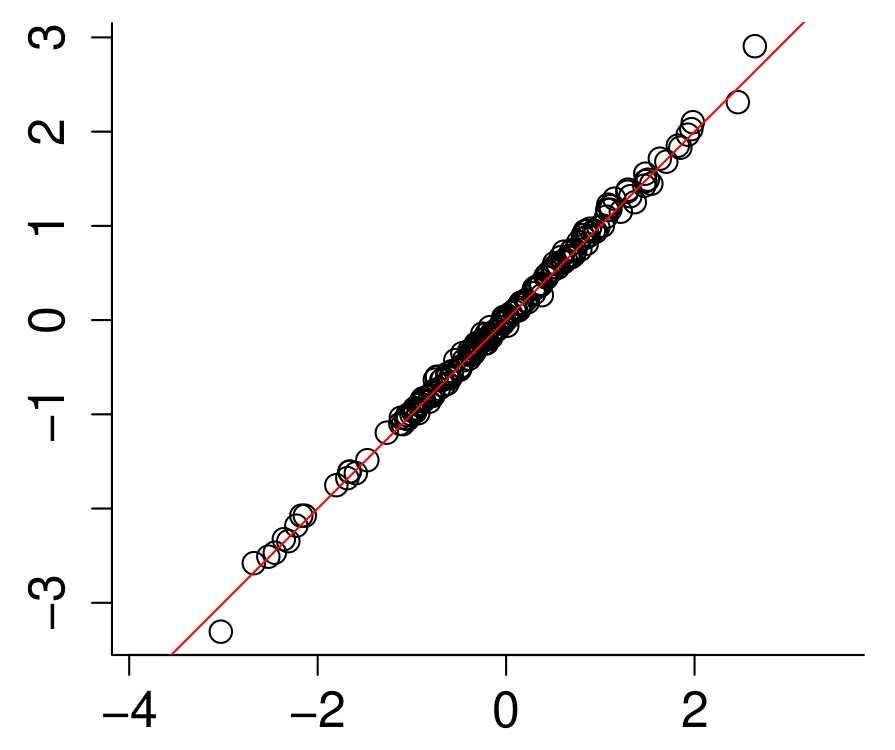}
		\caption{$b_i$ parameters.}
	\end{subfigure}
	~
	\begin{subfigure}[b]{0.25\textwidth}
		\includegraphics[width=\textwidth]{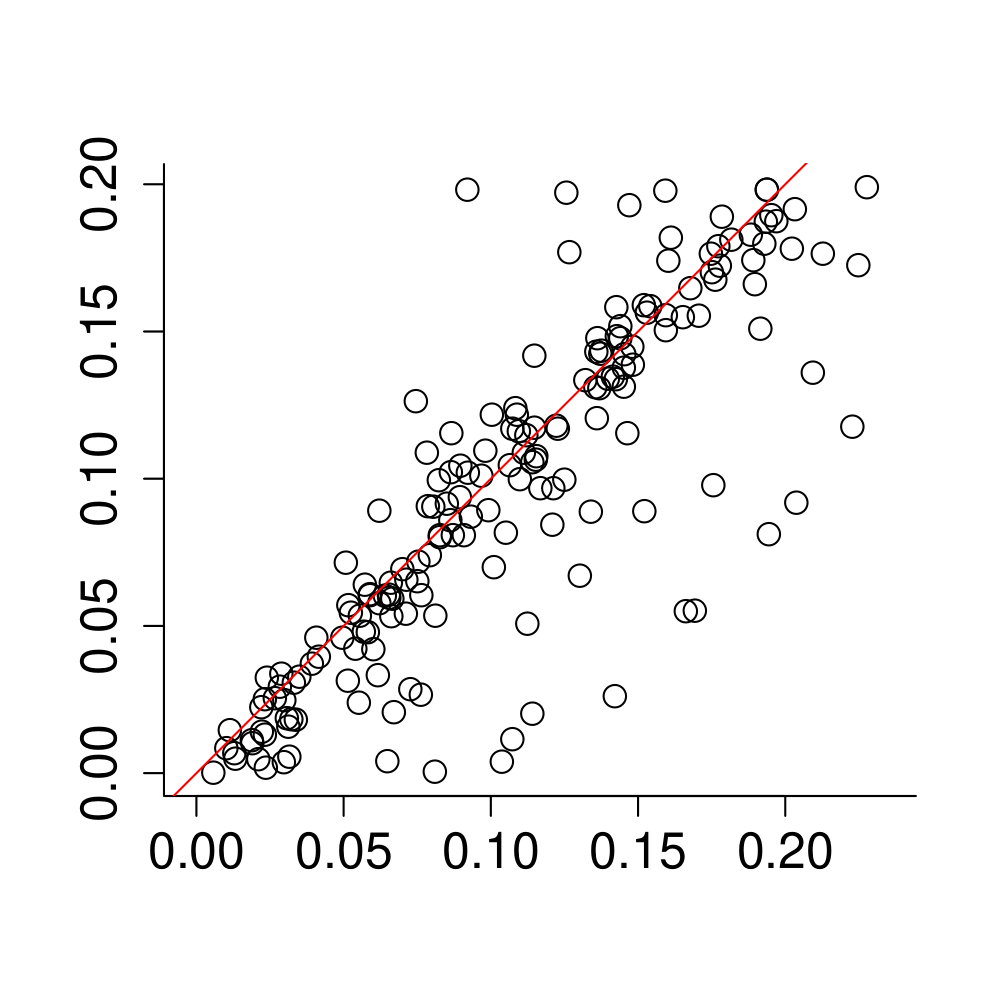}
		\caption{$c_i$ parameters.}
	\end{subfigure}
	~
	\begin{subfigure}[b]{0.25\textwidth}
		\includegraphics[width=\textwidth]{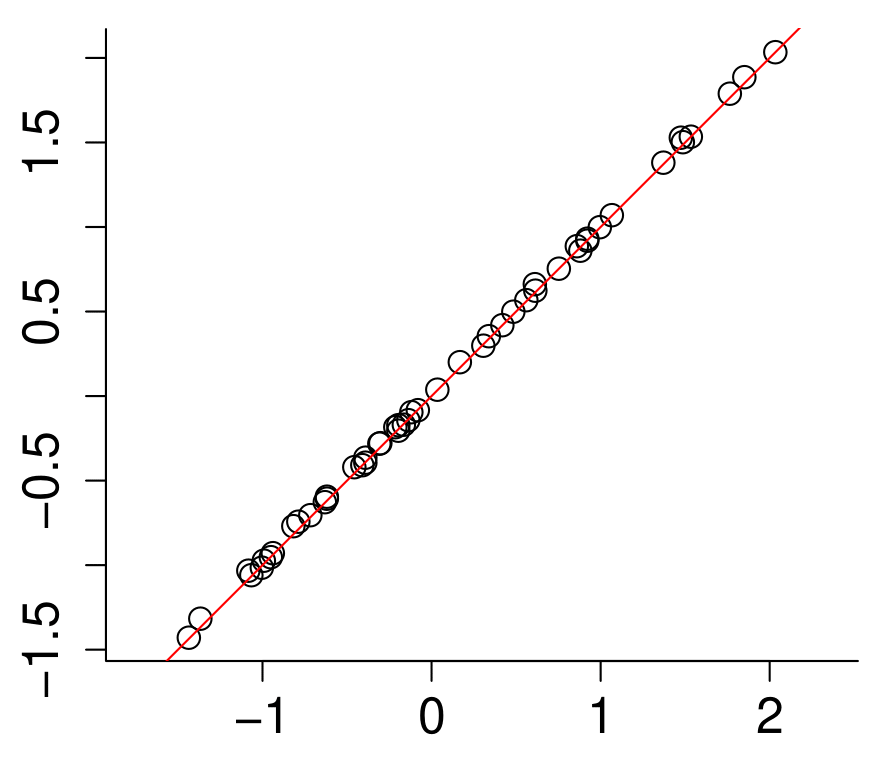}
		\caption{$b_{im}$ parameters.}
	\end{subfigure}
	\caption{True value (x-axis) versus posterior mean (y-axis) for simulation 8.}
	\label{graf:Completo}
\end{figure}

All $\bm{\lambda}$ parameters are well recovered (Table \ref{tab:Completo}). Figures \ref{graf:Simulation9_10k} and \ref{graf:Completo} (Appendix C) show good recover of all latent trait and item parameters. The pseudo parameters were middling recovered for some items, but this behavior is expected for these parameters, specially under such a highly structured model.

\begin{table}[h!]
	\centering
	\begin{tabular}{cccccc}\hline
		$\lambda$   & Mean & StandDev & $CI_{0.025}$& $CI_{0.975}$& CI Range\\ \hline
		0.95& 0.951& 0.002& 0.947& 0.956 &0.009\\
		0.90& 0.897& 0.003& 0.890& 0.902& 0.012\\
		0.85& 0.850& 0.004& 0.842& 0.857& 0.014\\
		0.80& 0.800& 0.004& 0.791& 0.808& 0.018\\
		0.75& 0.742& 0.006& 0.731& 0.754& 0.023\\\hline
	\end{tabular}
	\caption{Posterior statistics for $\bm{\lambda}$ in simulation 8.}
	\label{tab:Completo}
\end{table}

\section{Real data analysis - Enem}\label{realanalysis}

We apply the proposed methodology to a dataset from the 2017 High School National Exam (Enem) from Brazil. The exam is annually applied to high school students and is organized by the National Institute of Educational Studies and Researches Anísio Teixeira (Inep) from the Ministry of Education (MEC). It aims at assessing students who are concluding or have already concluded high school in the previous years and is used in the admission processes of universities around the country. Enem is composed of five sub-exams: Humanities (H), Natural Sciences (NS), Languages (Lang) and Maths (MT), each with 45 dichotomous unidimensional items, and an Essay. The first 45 items are related to Humanities, items 46 to 90 are related to Natural Sciences, items 91 to 135 are related to Language, items 136 to 180 are related to Mathematics, and, finally, items 181 to 185 are related to the Essay.

In order to obtain the subjects' final scores, Inep analyzes each of the four sub-exams with dichotomous items using the 3-parameter logistic IRT model. The Essay is corrected in a classical way, in which two referees independently grade the Essay over five matters, each one with grade 0, 20, 40, 60, 80, 100, 120, 140, 160, 180 or 200. The five grades are summed for each referee and the final score is the average of both scores. This means that the Essay is not in the same scale of the other four sub-exams, i.e., it is not under a normal scale with fixed mean and variance. The student's final score is the average of the five scores. This approach could be improved since it averages scores with different scales and non-trivial correlation structures.

It is reasonable to expect a multidimensional behavior for some items, since, for instance, some Mathematics items require more interpretation skills or some Natural Science items require some Mathematics skills. Moreover, it is expected to find a positive correlation among a same person's abilities. In fact, this is evident when computing the correlation matrix for scores obtained from Enem's grading system (for a random sample of 44,000 students - see Table \ref{tab:MatrixCorr1}).

\begin{table}[h!]
	\centering
	\begin{tabular}{lcccc}
		\hline
		& Humanities   & Natural Sciences  & Language   & Mathematics \\
		Natural Sciences    & 0.67 &      &      &      \\
		Language      & 0.62 & 0.75 &      &      \\
		Mathematics   & 0.63 & 0.64 & 0.60 &      \\
		Essay         & 0.48 & 0.53 & 0.54 & 0.47 \\ \hline
	\end{tabular}
	\caption{Correlation of the latent traits in the Enem 2017 exam.}\label{tab:MatrixCorr1}
\end{table}

The results in Table \ref{tab:MatrixCorr1} show a significant correlation among Enem's estimated abilities. Notice that the scores were obtained from an analysis in which latent traits were estimated separately for each sub-exam and, therefore, without considering any relation between specific skills. A more coherent statistical analysis requires the relationship among the specific abilities and the definition of the overall ability to be properly considered in one joint model. In the same way, the possible multidimensional structure of some items should also be considered. Moreover, no source of uncertainty should be ignored in the estimation of traits and their relationships.

We consider a random sample of 44,000 students from all over the country. There are four types of exams, differing only by item's ordering and we use the items ordering according to the Blue version of the exam. For the Languages exam, students can choose between English or Spanish as a foreign language but the sample was chosen only among students who chose the English exam. The first five items in the Language exam are related to the chosen foreign language. The sample contains only students who completed the five exams and that were not disqualified in any of them. We subjectively chose some multidimensional items based on textual and mathematics complexity (see Table \ref{tab:PostMeanMultiItems}). After a pilot analysis, some items were estimated as having negative discrimination and were excluded from the analysis (10, 23, 25, 30, 36, 65, 123, 143 and 157). This happens when the characteristic curve is not monotonically increasing.

Table \ref{tab:PostLambdaEnem} presents the posterior statistics for factor weights $\bm{\lambda}$, showing that the highest value was related to the Humanities ($\lambda_{H}$ = 0.98) exam, followed by Natural Sciences (0.964) and the lowest was Essay (0.686). These results show that, as expected, all the sub-exams are highly correlated. Notice that, if the exam as a whole was unidimensional, all weights $\bm{\lambda}$ would be close to one. Moreover, if the sub-exams were each unidimensional, but uncorrelated, as it is assumed by Inep to compute the scores, all the weights $\bm{\lambda}$ would be close to zero. Therefore, it is clear that a simple average over the five scores is not the most reasonable way to estimate the general ability, in the same way that it is not reasonable to score subjects only over the number of correct responses.

We highlight the results for item 86 which is in the Natural Sciences exam but was estimated to have a high coefficient for the Maths ability.

\begin{table}[h!]
	\centering
	\begin{tabular}{lcccccc}\hline
		Exam   & Mean & StandDev & $CI_{0.025}$& $CI_{0.975}$ & CI Range  \\\hline
		Humanities&0.980 &0.001& 0.979& 0.982& 0.004\\
		Natural Sciences&0.964& 0.001& 0.962& 0.967& 0.005\\
		Language &0.936& 0.001& 0.934& 0.939& 0.005\\
		Mathematics &0.892& 0.002& 0.886& 0.896& 0.010\\
		Essay &0.686& 0.003& 0.680& 0.692& 0.012\\\hline
	\end{tabular}
	\caption{Posterior values for $\bm{\lambda}$ - Enem.}
	\label{tab:PostLambdaEnem}
\end{table}

\begin{table}[h!]
	\centering
	\begin{tabular}{cccccccc}\hline
		Item &$a_{H}$  & $a_{NS}$  & $a_{Lang}$ & $a_{MT}$ & $a_{Essay}$& b&c   \\\hline
		18	&	0.91	&		&	0.31	&		&		&	1.44	&	0.24	\\
		43	&	0.63	&		&	0.40	&		&		&	1.52	&	0.21	\\
		84	&		&	1.00	&		&	0.73	&		&	3.35	&	0.08	\\
		\textbf{86}	&		&	\textbf{0.65}	&		&	\textbf{1.07}	&		&	4.66	&	0.09	\\
		156	&		&		&	0.24	&	2.23	&		&	5.28	&	0.21	\\
		181	&		&		&	0.20	&		&	1.31	&		&		\\\hline
	\end{tabular}
	\caption{Posterior mean for item parameters for the multidimensional item - Enem.}
	\label{tab:PostMeanMultiItems}
\end{table}

 \begin{table}[h!]
	\centering
	\begin{tabular}{lcccc}
		\hline
		& Humanities   & Natural Sciences  & Language   & Mathematics   \\
		Natural Sciences       & 0.99&     &     &     \\
		Language               & 0.97& 0.96&     &      \\
		Mathematics            & 0.95& 0.95& 0.92&      \\
		Essay                  & 0.73& 0.73& 0.70& 0.69 \\ \hline
	\end{tabular}
	\caption{Correlations among posterior latent trait means - Enem.}\label{tab:MatrixCorr2}
\end{table}

\section{Conclusions}

This paper proposed a general methodology to perform statistical analysis of item response data in the presence of a hierarchical multidimensional structure for the latent traits. Considerable limitations in terms of modeling and inference from existing solutions are discussed and overcome in the proposed methodology.

The proposed IRT model is specified in a way that all the latent traits are in the same fixed scale so that model identification is guaranteed and interpretation is not lost. We studied computational aspects and carefully devised an MCMC algorithm to make reliable inference.

Simulation studies were performed to investigate the efficiency of the proposed MCMC algorithm under different sample sizes. Moreover, we studied the impact of not considering the hierarchical structure when this exists and vice-versa. We showed that the model is capable of detecting whether the test is unidimensional, multidimensional with uncorrelated latent traits or multidimensional with correlated latent traits.

Finally, we applied the proposed methodology to a real data set from the High School National Exam (Enem) from Brazil, which is the most important education exam in the country. Differently from the currently used methodology to score the Enem exam, we considered a hierarchical structure for the latent traits, allowing us to estimate a more robust and informative general latent trait. We also considered some multidimensional items for which such feature was confirmed. Finally, the Essay's score was estimated via IRT together with the other sub-exams, which also contributed to the quality of the estimated general latent trait.

A possible direction for future work could be to consider more complex structures for the latent traits like, for example, allowing one trait to depend on more than one trait from its upper level, which should be useful in psychological applications.



\bibliographystyle{apa}
\bibliography{bibliografia}

\newpage

\section*{Appendix A - Proof of Proposition \ref{prop1}}

\begin{proof}
	
	The specification in \textbf{A3} implies that $\bm{\theta}^{(K)}$ is independent of $\bm{\lambda}^{(K)}$, therefore
	\begin{equation}\label{eq:K1}
	\begin{split}
	\pi(\theta_{qj}^{(K-1)}) &= \int \int \pi(\theta_{qj}^{(K-1)},\theta_{qj}^{(K)},\lambda_{q}^{(K)} ) \quad d\theta_{qj}^{(K)}d\lambda_{q}^{(K)} \\
	&= \int \int \pi(\theta_{qj}^{(K-1)}|\theta_{qj}^{(K)},\lambda_{q}^{(K)} )\pi(\theta_{qj}^{(K)},\lambda_{q}^{(K)}) \quad d\theta_{qj}^{(K)}d\lambda_{q}^{(K)} \\
	&= \int \int \pi(\theta_{qj}^{(K-1)}|\theta_{qj}^{(K)},\lambda_{q}^{(K)} )\pi(\theta_{qj}^{(K)})\pi(\lambda_{q}^{(K)}) \quad d\theta_{qj}^{(K)}d\lambda_{q}^{(K)} \\
	&= \int_{\lambda}\biggl[ \int_{\theta} \pi(\theta_{qj}^{(K-1)}|\theta_{qj}^{(K)},\lambda_{q}^{(K)} )\pi(\theta_{qj}^{(K)})d\theta_{qj}^{(K)} \biggr] \pi(\lambda_{q}^{(K)})\quad d\lambda_{q}^{(K)}\\
	&= \int_{\lambda} \phi(\theta^{(K-1)}_{qj}) \pi(\lambda_{q}^{(K)}) \quad d\lambda_{q}^{(K)}\\	
	&= \phi(\theta^{(K-1)}_{qj}) \int_{\lambda}  \pi(\lambda_{q}^{(K)}) \quad d\lambda_{q}^{(K)}\\	
	&= \phi(\theta^{(K-1)}_{qj}).
	\end{split}
	\end{equation}
	For $k = K-2$, it follows that
	\begin{equation}\label{eq:K2}
	\begin{split}
	\pi(\theta_{qj}^{(K-2)}) &= \int \int \pi(\theta_{qj}^{(K-2)}|\theta_{q'j}^{(K-2)},\lambda_{q}^{(K-2)} )\pi(\theta_{q'j}^{(K-1)},\lambda_{q}^{(K-1)}) \quad d\theta_{q'j}^{(K-1)}d\lambda_{q}^{(K-1)}
	\end{split}
	\end{equation}
	where
	\begin{equation}\label{eq:K2cont}
	\begin{split}
	\pi(\theta_{q'j}^{(K-1)},\lambda_{q}^{(K-1)}) &= \int \int \pi(\theta_{q'j}^{(K-1)}|\theta_{q''j}^{(K)},\lambda_{q'}^{(K)} )\pi(\lambda_{qj}^{(K-1)}|\theta_{q''j}^{(K)},\lambda_{q'}^{(K)} )\pi(\lambda_{q'}^{(K)})\pi(\theta_{q'j}^{(K)}) d\lambda_{q'}^{(K)} d\theta_{q'j}^{(K)}\\
	&= \pi(\lambda_q^{(K-1)})(\theta_{q'j}^{(K-1)}).
	\end{split}
	\end{equation}
	For $k \leq K-3$, the proof is analogous and devised recursively such that the result in (\ref{eq:K2cont}) for step $k$ is used on the step for $k - 1$.
\end{proof}

\section*{Appendix B - MCMC details}

We present the algorithms to simulate from each of the full conditional distributions in the proposed Gibbs sampling algorithm from Section \ref{secmcmc}. It is assumed that all the $J$ individuals response all the $I$ items in the test. Adaptations are straightforward if that is not the case by simply ignoring the respective likelihood terms.

\subsection*{Sampling $(\bm{Z}, \bm{X})$}

The pairs ($Z_{ij}, X_{ij}$) are conditionally independent for all $(i,j)$ and
\begin{equation}\label{CondZX}
(Z_{ij}, X_{ij})|. =\left\{
\begin{array}{ll}
\phi(x_{ij} - m_{ij})\I(Z_{ij}=0)\I(X_{ij}<0),&\quad\mbox{if}\quad Y_{ij}=0,\\
w_{ij}\I(Z_{ij}=1)\I(X_{ij}=0)+(1-w_{ij})\frac{\phi(x_{ij} - m_{ij})}{\Phi( m_{ij})}\I(Z_{ij}=0)\I(X_{ij} > 0)&\quad \mbox{if}\quad Y_{ij}=1,
\end{array}
\right.
\end{equation}
where $m_{ij} = \textbf{a}_i\bm{\theta}^{(1)}_j-b_i$ and $w_{ij} = \frac{c_i}{c_i+(1-c_i)\Phi(m_{ij})}$.

This means that, if $Y_{ij} = 0, Z_{ij}$ is a point mass at zero and $X_{ij}$ is a N($m_ij$, 1) truncated to be negative. On the other hand, if $Y_{ij}$ = 1, $Z_ij \sim Ber(w_{ij})$ and $X_{ij}$ is a point-mass at 0 if $Z_{ij}$ = 1 and is $N(m_{ij}, 1)$ truncated to be positive if $Z_{ij}$ = 0.

\subsection*{Sampling $\bm{\theta}$}

Sampling the whole vector $\bm{\theta}$ jointly is possible due to the conditional independence of the respective full conditional distributions among different subjects.
The joint full conditional distribution of all $\bm{\theta}_j$ can be factorized as
\begin{equation}\label{FullConditionalThetaConj2}
\pi(\bm{\theta}_j^{(1)},\dots,\bm{\theta}_j^{(K)}|.) \propto  \pi(\bm{\theta}_j^{(K)}|.)\pi(\bm{\theta}^{(K-1)}|\bm{\theta}_j^{(K)}, .)\dots\pi(\bm{\theta}_j^{(1)}|\bm{\theta}_j^{(2)},\dots,\bm{\theta}_j^{(K)}, .),
\end{equation}
where, for $k = 1, \dots K$,
\begin{equation}
\pi(\bm{\theta}^{(k)}|.)\sim N_{Q_k}(\mu_j^{(k)},\Lambda^{(k)}),
\end{equation}
\begin{equation}
\Lambda^{(k)} = ({\Sigma^{(k+1)}}^{-1}+{\lambda^{(k+1)}}^T{\Sigma^{(k+1)}}^{-1}\lambda^{(k+1)} - {\lambda^{(k)}}^T{\Sigma^{(k)}}^{-1}\Lambda^{(k-1)}{\Sigma^{(k)}}^{-1}\lambda^{(k)})^{-1},
\end{equation}
and
\begin{equation}
\mu_j^{(k)} = \Lambda^{(k)}({\Sigma^{(k+1)}}^{-1}\lambda^{(k+1)}\theta^{(k+1)} + \biggl[\prod_{i=2}^{k} {\lambda^{(i)}}'{\Sigma^{(i)}}^{-1}\Lambda^{(i-1)}\biggr]a'(X_j+b)),
\end{equation}
where $\Sigma^{(k)}=diag\left( 1-\sum_{q'=1}^{Q_k}(\lambda_{1q'}^{(k)})^2,\ldots,1-\sum_{q'=1}^{Q_k}(\lambda_{Q_{k-1}q'}^{(k)})^2\right)$, $\lambda^{(K+1)}=0$, $\theta^{(K+1)} = 0$ and ${\Sigma^{(K+1)}}=1$. Also, for each $j$, if $z_{ij} = 0$, $(\textbf{a}_i, b_i) = 0$.

We sample from (\ref{FullConditionalThetaConj2}) recursively from $\bm{\theta}_{j}^{(K)}$ to $\bm{\theta}_{1}^{(1)}$.

\subsection*{Sampling $\bm{c}$}

Assuming independent $Beta(\alpha,\beta)$ priors, all the $c_i$ parameters are conditionally independent with
\begin{equation}
\pi(c_i|.) \sim Beta\bigg(\sum_{j=1}^JZ_{ij} + \alpha, J - \sum_{j=1}^JZ_{ij} + \beta\bigg).
\end{equation}

\subsection*{Sampling $(\bm{a},\bm{b})$}

Define $\textbf{X}_i = (X_{i1}, \dots, X_{iJ})'$ as a $J\times 1$ vector and $\textbf{C}$ as the $(Q_1+1)\times J$ matrix with $q$-th row given by $(\theta_{q1},\ldots,\theta_{qJ})$ and the last row being all equal to -1. The vectors $(\bm{a}_i,b_i)$ are conditionally independent among items with
\begin{equation}\label{CondCompletaAB}
(\textbf{a}_i, b_i)|. \sim N(\bm{\mu}^*_{\eta_i}, \bm{\Lambda}^*_{\eta}),
\end{equation}
where
\begin{equation}\label{eq:VarAB}
\bm{\Lambda}^*_{\eta} = (\bm{\Lambda}_{\eta}^{-1} + \textbf{C}\textbf{C}')^{-1}, \quad and \quad \bm{\mu}^*_{\eta_i} = \bm{\Lambda}^*_{\eta}(\bm{\Lambda_{\eta}}^{-1}\bm{\mu}_{\eta_i} + \bm{C}\textbf{X}_i).
\end{equation}

\subsection*{Sampling $\bm{\lambda}$}

The non-null coordinates of all the $\bm{\lambda}_{q}^{(k)}$ are mutually independent and each one has a full conditional density proportional to
\begin{equation}\label{eq:CondLambda}
\pi(\lambda_{qq'}^{(k)}|.) \propto \biggl[\prod_j \frac{1}{\tau_q^{(k)}\sqrt{2\pi}} exp\biggl\{-\frac{1}{2}\biggl( \frac{\theta_{qj}^{(k-1)} - \lambda_{qq'}^{(k)}\bm{\theta}_{q'j}^{(k)}}{\tau_q^{(k)}}\biggl)^2 \biggr\} \biggr] \I(0 < \lambda_{qq'}^{(k)} <1),
\end{equation}
where $\tau_q^{(k)} = \sqrt{1- {\lambda_{qq'}^{(k)}}^2}$.

We sample from this density using a Metropolis Hastings step with a Gaussian random walk proposal, properly tuned to optimise convergence speed (see \cite{roberts1997weak}).

\subsection*{Sampling $\bm{X}^G$}

The auxiliary variables $(X_{ij})^G$ are conditionally independent for all $(i,j)$ with
\begin{equation}\label{CondCompletaXGraded}
(X^G_{ij}| \bm{\theta}, \textbf{a}, \textbf{b}_G, \textbf{Y}) \sim
N(\textbf{a}^G_i\bm{\theta}^{(1)}_j,1) \I (b_{i,m-1} < X^G_{ij} < b_{i,m}),\mbox{if}\quad Y_{ij} = m,
\end{equation}

\subsection*{Sampling $\bm{b}^G$}

Due to the typically high number of $X^G_{ij}$ variables, the full conditional distribution of the location parameters $\bm{b}^G$ will be truncated to very small intervals, leading to very poor mixing of the chain. In order to overcome this issue, we adopt a collapsed Gibbs sampler \citep[see][]{liu1994collapsed} strategy to sample those parameters by integrating out $\bm{X}^G_{ij}$. Also, to avoid the difficulties in devising an efficient MH proposal due to the ordering constraint of the $\bm{b}^G$ parameters, we consider the following reparametrization:
\begin{align*}\label{eq:reparaB}
b_{i1}^{t*} &= b^{t}_{i1}\\
b_{i2}^{t*} &= b^{t}_{i2} - b^{t}_{i1}\\
\vdots \\
b_{i,M_i-1}^{t*} &= b^t_{i,M_i-1} - b^t_{i,M_i-2}
\end{align*}
where  $b_{i0}^{*} = b_{i0} = -\infty$ and $b_{i,M_i}^{*} = b_{i,M_i} = \infty$

We perform two different update steps for the reparametrized location parameters. The first one (Algorithm 1) only updates $b_{i,1}^*$ using a (properly tuned) Gaussian random walk proposal. This implies in a translation of all the original parameters, preserving their differences. The second one (Algorithm 2) updates the $M_i-2$ difference parameters $b_{i,2:M-1}^*$ using a (properly tuned) Gaussian random walk proposal. This means that $b_{i,1}$ is preserved and the respective differences between the $b_{i,}$'s are updated. The two update steps are as follows.

	{\bf Algorithm 1}
	\begin{algorithmic}[1]
		\Statex
		 On iteration $t$, with current value $\bm{b}_i$:
		\For{graded item $i$:}
		\State simulate $\textbf{b}^{c}_{i,1:M_i-1} \sim N(\textbf{b}_{i,1:M_i-1},\Sigma)$,
		\State accept proposal with probability $\alpha(\textbf{b}_i,\textbf{b}^c_i)= 1 \wedge \frac{p(\textbf{b}^{c}_{i,1:M_i-1})}{p(\textbf{b}_{i,1:M_i-1})}$.
		\EndFor
	\end{algorithmic}

	{\bf Algorithm 2}
	\begin{algorithmic}[1]
		\Statex
		On iteration $t$, with current value $\bm{b}_{i}^*$:
		\For{graded item $i$:}
		\State simulate $b_{i,2:M_i-1}^{*c} \sim N_{(M_i-2)}(\textbf{b}_{i}^* ,\bm{\Sigma})\I(b_{2:M_i-1}>0)$,
		\State accept proposal with probability $\alpha(\textbf{b}_{i}^*,\textbf{b}^{*c}_i)= min \Big\{1,\frac{p(\textbf{b}^{*c}_{i,2:M_i-1})}{p(\textbf{b}_{i,2:M_i-1}^*)}\Big\}$.
		\EndFor
	\end{algorithmic}
where $p(.)$ is given the product of (\ref{LikeGraded}) and the prior density of $\bm{b}_i$.

\subsection*{Sampling $\bm{a}^G$}

Parameters $\textbf{a}_{i}^G$ are conditionally independent among items with full conditional distribution given by
\begin{equation}\label{CondCompletaABGrade}
(\textbf{a}^G_i)|. \sim N(\bm{\mu}^*_{a_i}, \bm{\Lambda}^*_{a}),
\end{equation}
where
\begin{equation}\label{eq:VarA}
\bm{\Lambda}^*_{a} = ({\bm{\Lambda}_{a}}^{-1} + \bm{\theta}^{(1)}(\bm{\theta}^{(1)\prime})^{-1}, \quad and \quad \bm{\mu}^*_{a_i} = \bm{\Lambda}^*_{a}({\bm{\Lambda}_{a}}^{-1}\bm{\mu}_{a} + \bm{\theta}^{(1)}\textbf{X}_i).
\end{equation}

\newpage

\section*{Appendix C - Further results from the simulations}

\begin{figure}[h!]
	\centering
	\begin{subfigure}[b]{0.25\textwidth}
		\includegraphics[width=\textwidth]{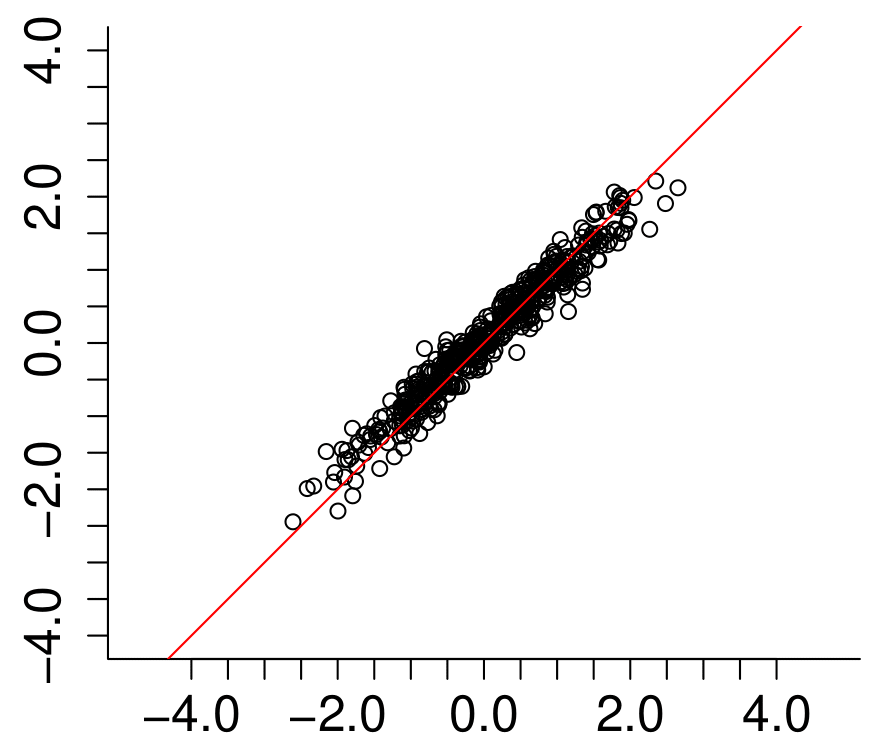}
		\caption{$\theta_1^{(1)}$}
	\end{subfigure}
	~
	\begin{subfigure}[b]{0.25\textwidth}
		\includegraphics[width=\textwidth]{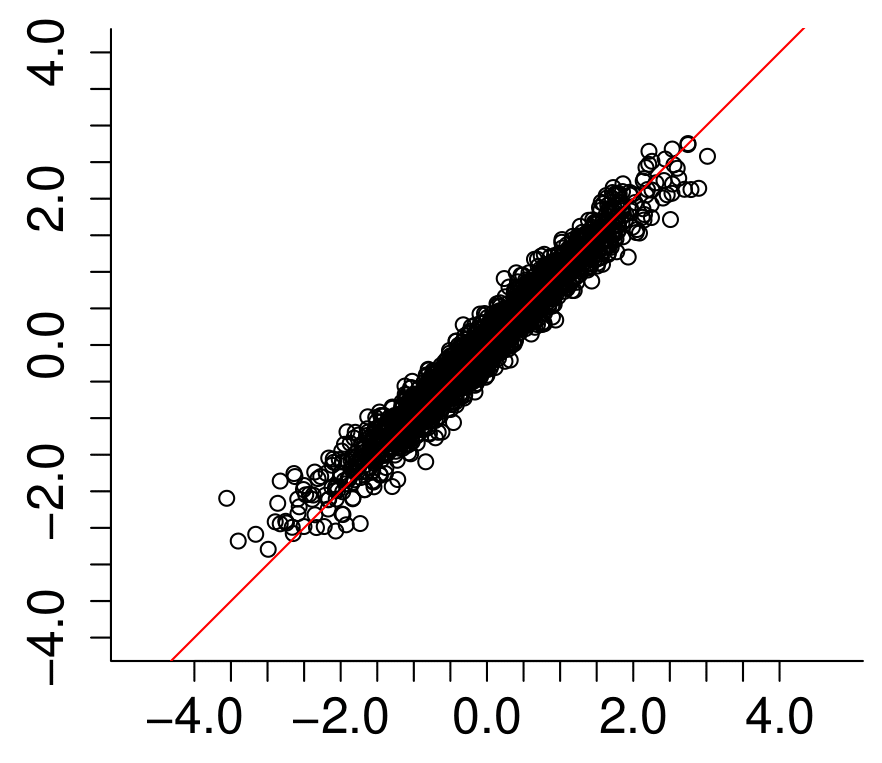}
		\caption{$\theta_1^{(1)}$}
	\end{subfigure}
	~
	\begin{subfigure}[b]{0.25\textwidth}
		\includegraphics[width=\textwidth]{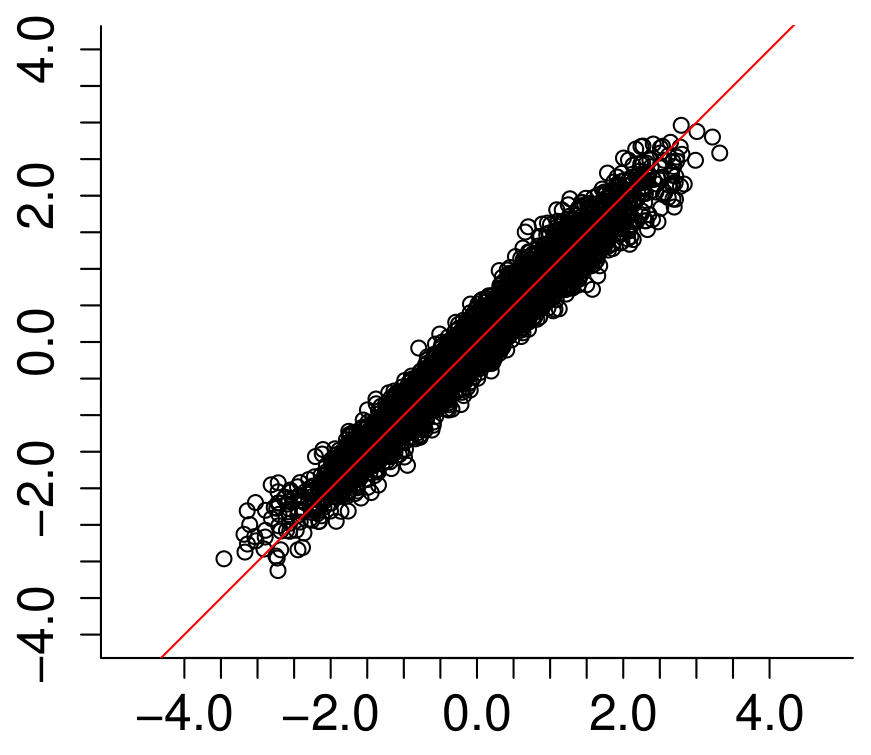}
		\caption{$\theta_1^{(1)}$}
	\end{subfigure}
	~ 
	\begin{subfigure}[b]{0.25\textwidth}
		\includegraphics[width=\textwidth]{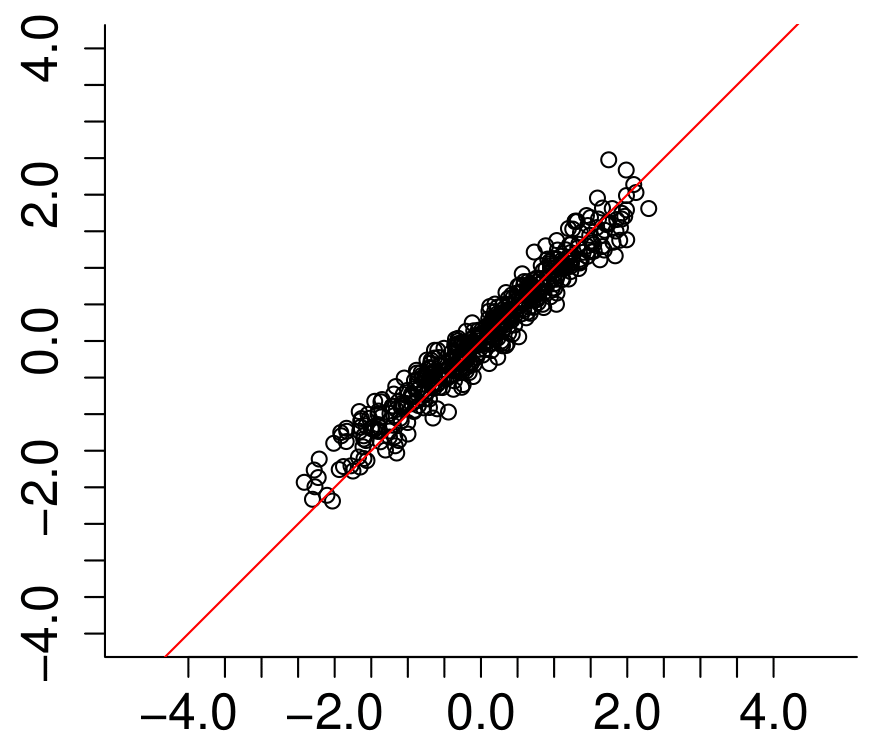}
		\caption{$\theta_2^{(1)}$}
	\end{subfigure}
	~
	\begin{subfigure}[b]{0.25\textwidth}
		\includegraphics[width=\textwidth]{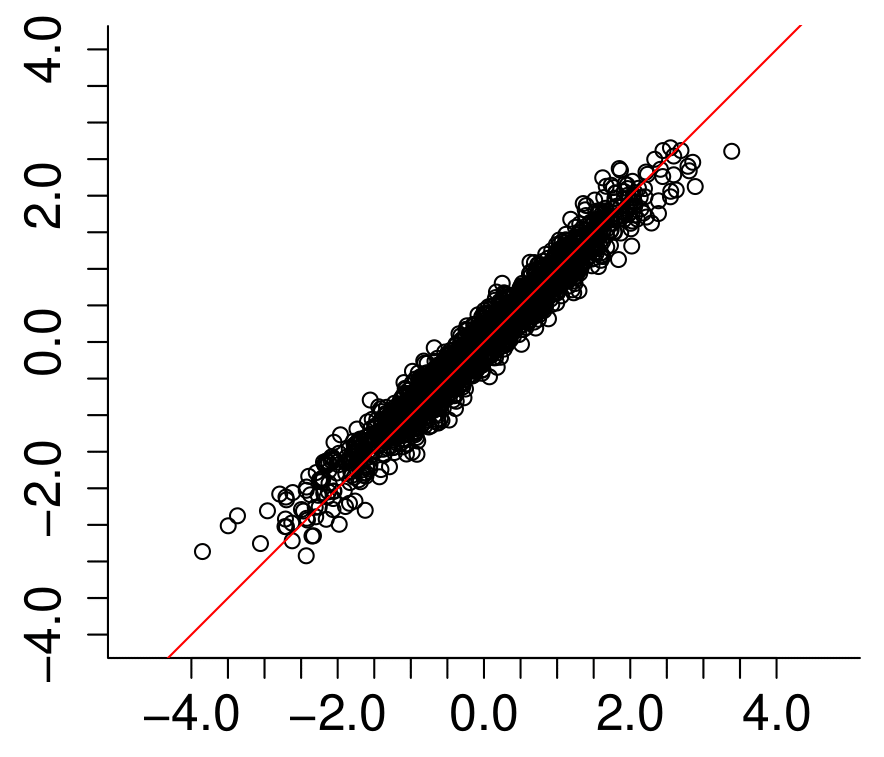}
		\caption{$\theta_2^{(1)}$}
	\end{subfigure}
	~
	\begin{subfigure}[b]{0.25\textwidth}
		\includegraphics[width=\textwidth]{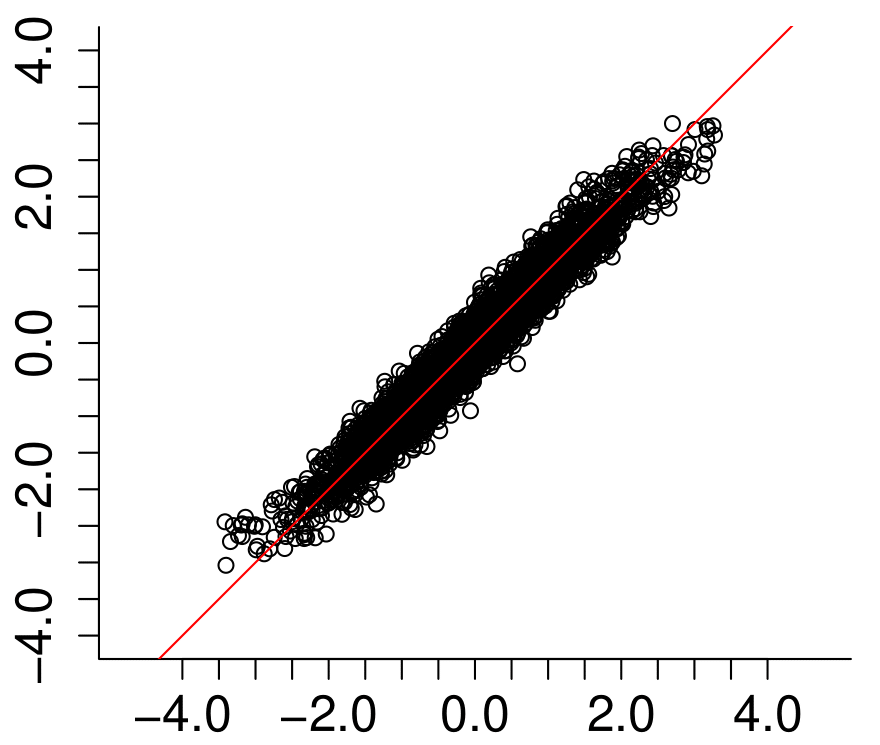}
		\caption{$\theta_2^{(1)}$}
	\end{subfigure} 	
	~ 
	\begin{subfigure}[b]{0.25\textwidth}
		\includegraphics[page=1,width=\textwidth]{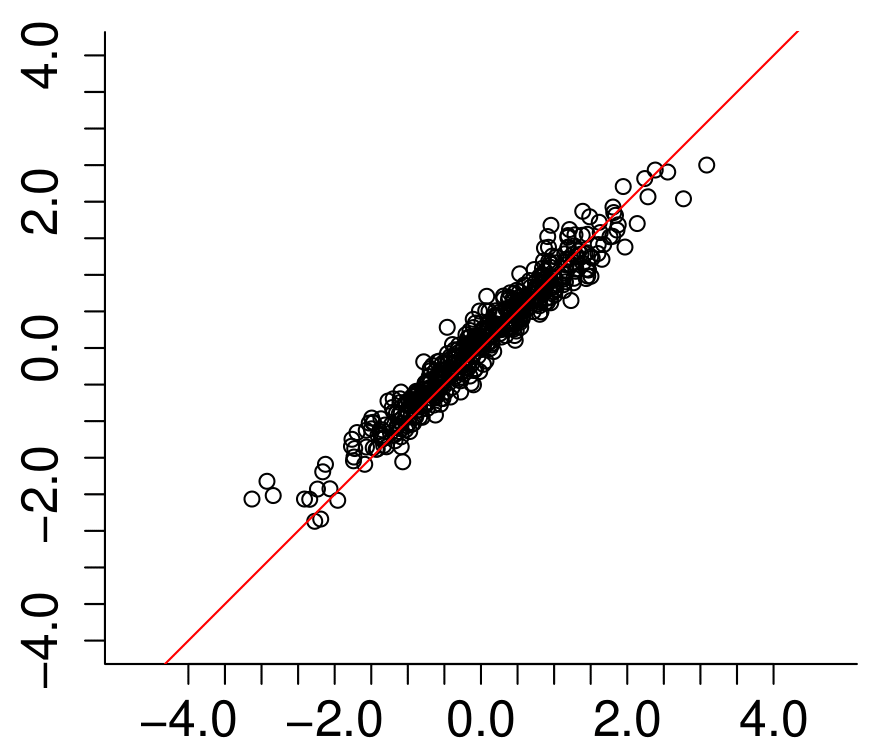}
		\caption{$\theta_3^{(1)}$}
	\end{subfigure}
	~
	\begin{subfigure}[b]{0.25\textwidth}
		\includegraphics[width=\textwidth]{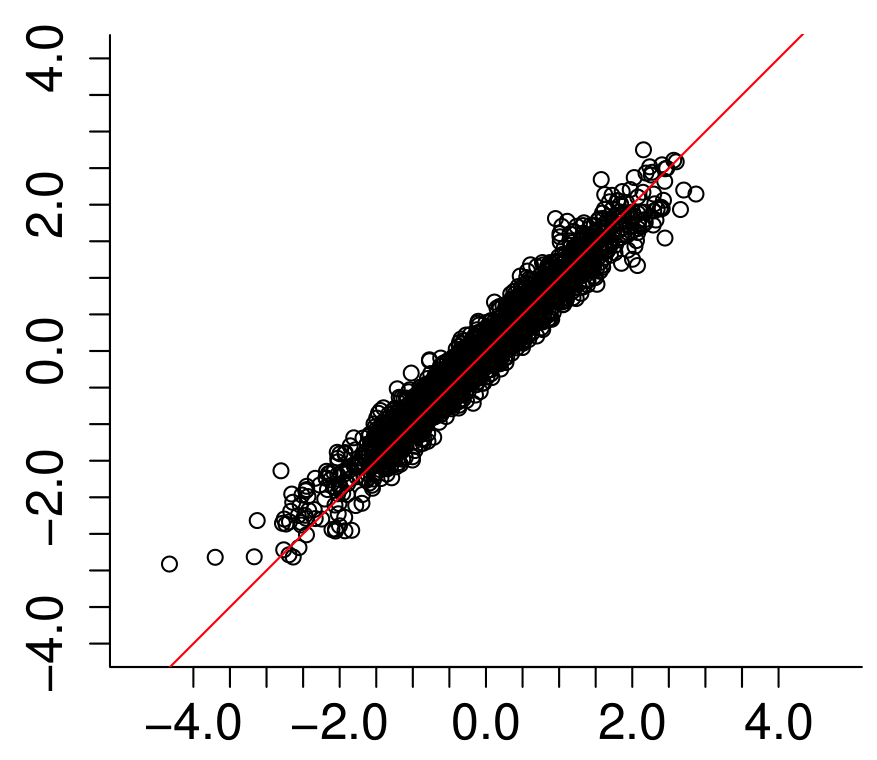}
		\caption{$\theta_3^{(1)}$}
	\end{subfigure}
	~
	\begin{subfigure}[b]{0.25\textwidth}
		\includegraphics[width=\textwidth]{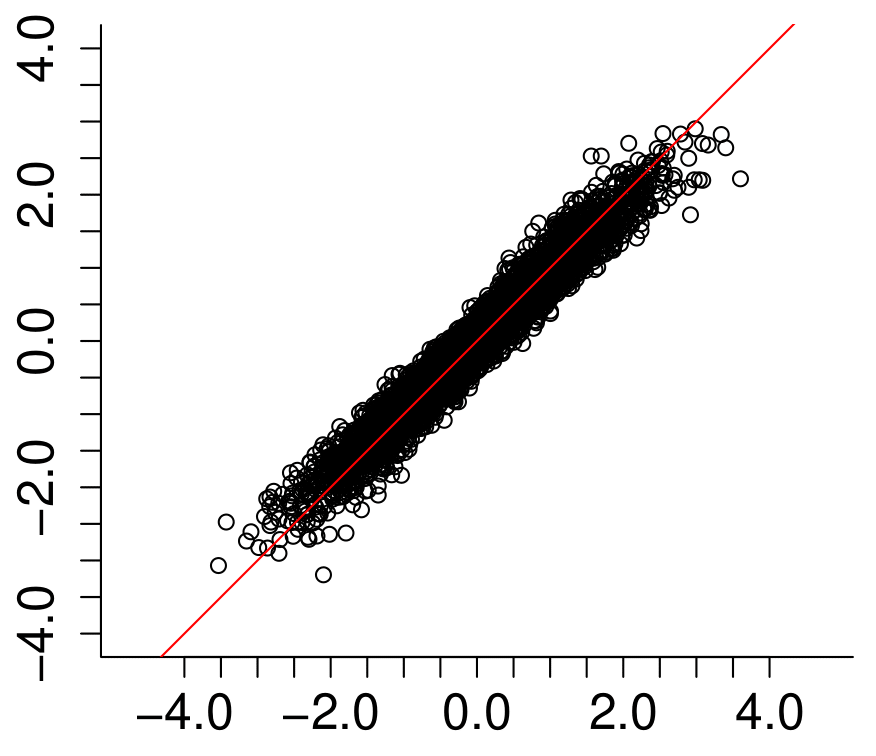}
		\caption{$\theta_3^{(1)}$}
	\end{subfigure}
	~
	\begin{subfigure}[b]{0.25\textwidth}
		\includegraphics[width=\textwidth]{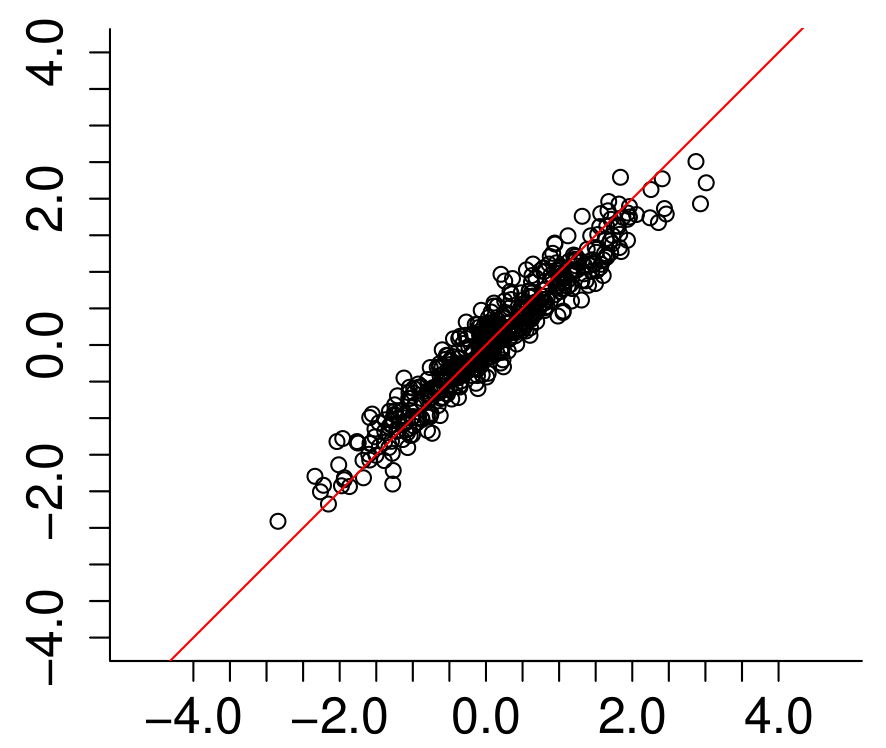}
		\caption{$\theta_4^{(1)}$}
	\end{subfigure}
	~
	\begin{subfigure}[b]{0.25\textwidth}
		\includegraphics[width=\textwidth]{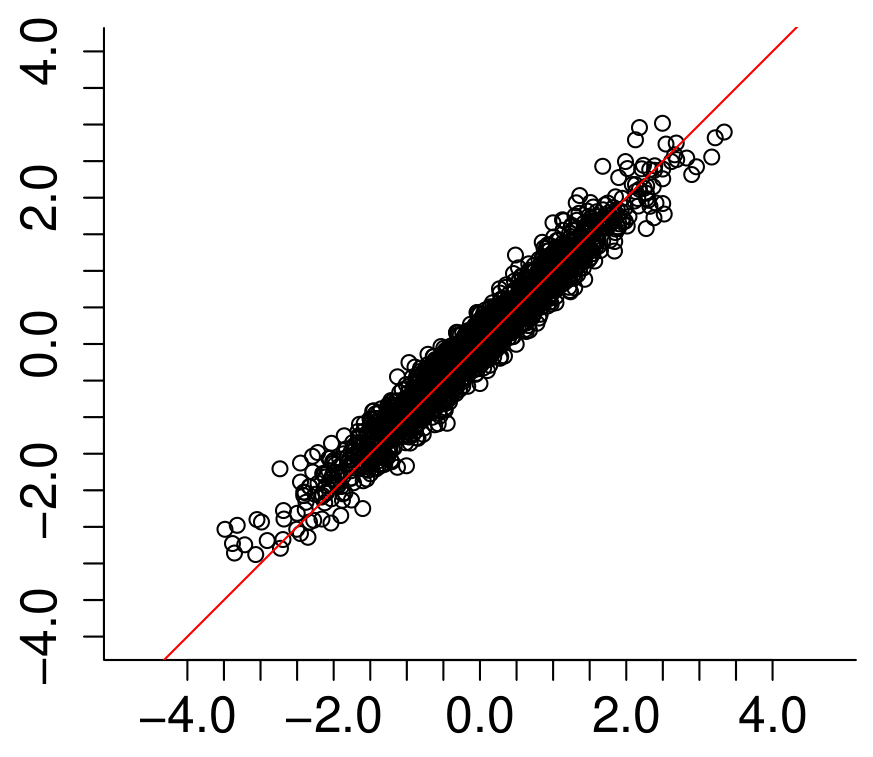}
		\caption{$\theta_4^{(1)}$}
	\end{subfigure}
	~
	\begin{subfigure}[b]{0.25\textwidth}
		\includegraphics[width=\textwidth]{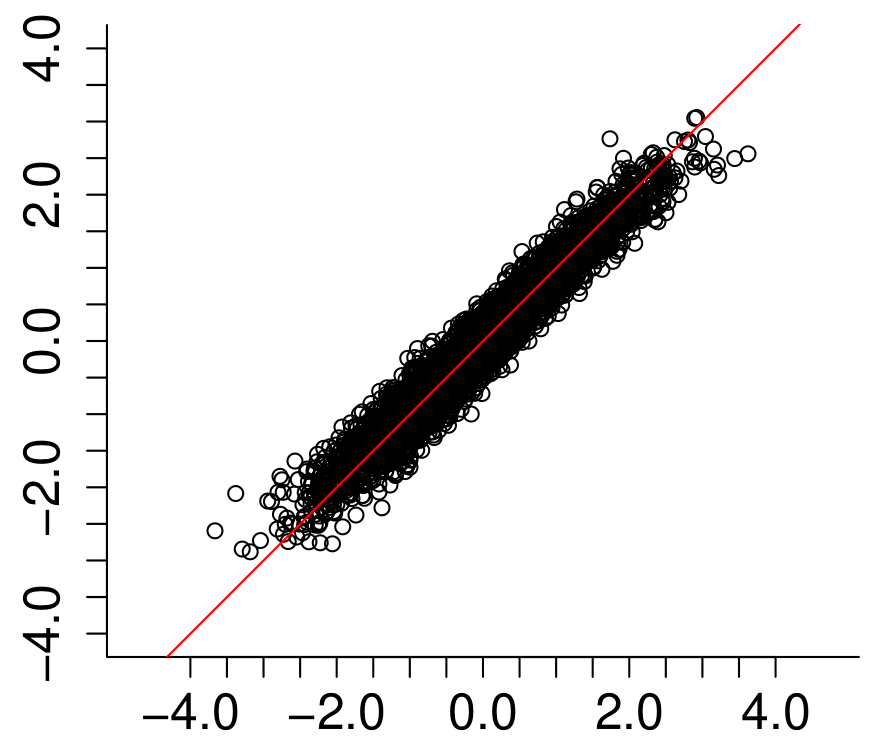}
		\caption{$\theta_4^{(1)}$}
	\end{subfigure}
	\caption{Posterior mean (y-axis) versus true (x-axis) values for 500 (left column), 2k (middle column) and 5k (right column) subjects (simulations 1-3).}
	\label{graf:Simulacao7_500_thetas}
\end{figure}

\begin{figure}[h!]
	\centering
	\begin{subfigure}[b]{0.24\textwidth}
		\includegraphics[width=\textwidth]{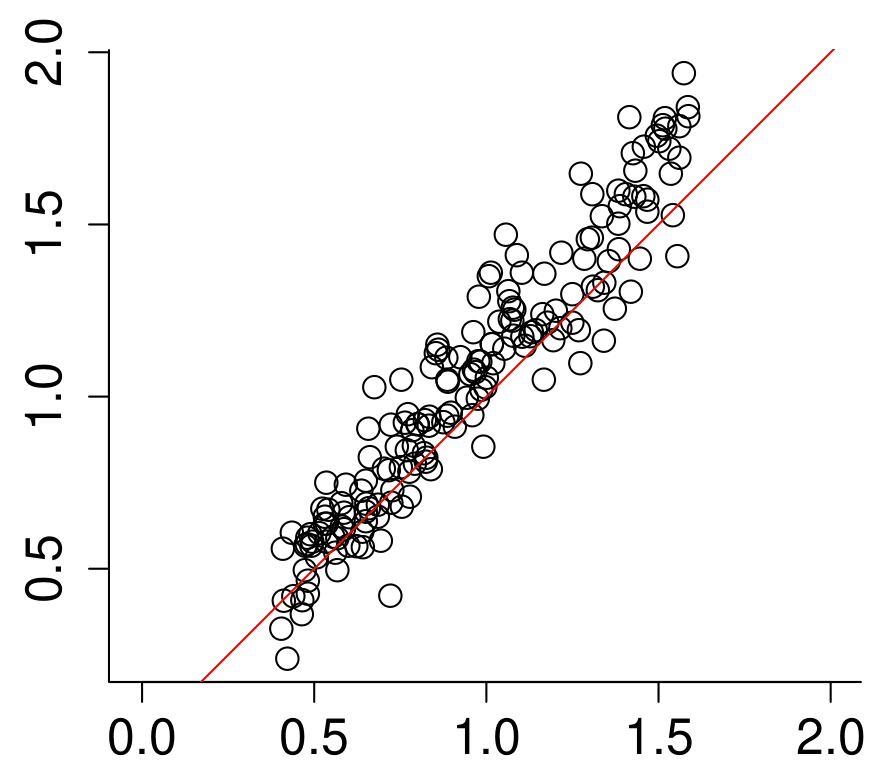}
		\caption{Discrimination - 500.}
	\end{subfigure}
	~
	\begin{subfigure}[b]{0.24\textwidth}
		\includegraphics[width=\textwidth]{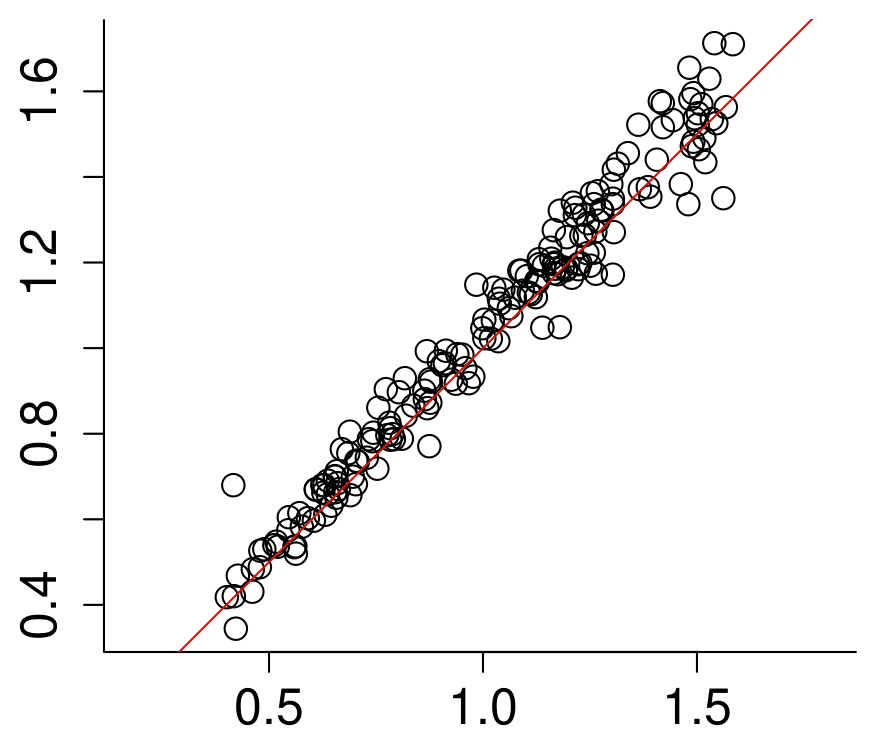}
		\caption{Discrimination - 2k.}
	\end{subfigure}
	~
	\begin{subfigure}[b]{0.24\textwidth}
		\includegraphics[width=\textwidth]{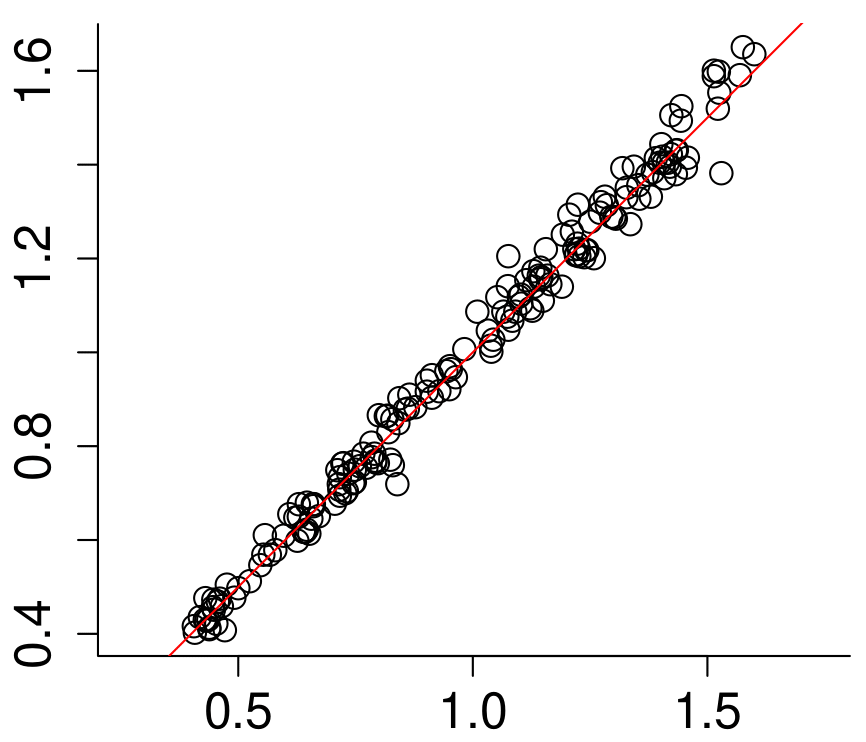}
		\caption{Discrimination - 5k.}
	\end{subfigure}
	~ 
	\begin{subfigure}[b]{0.24\textwidth}
		\includegraphics[width=\textwidth]{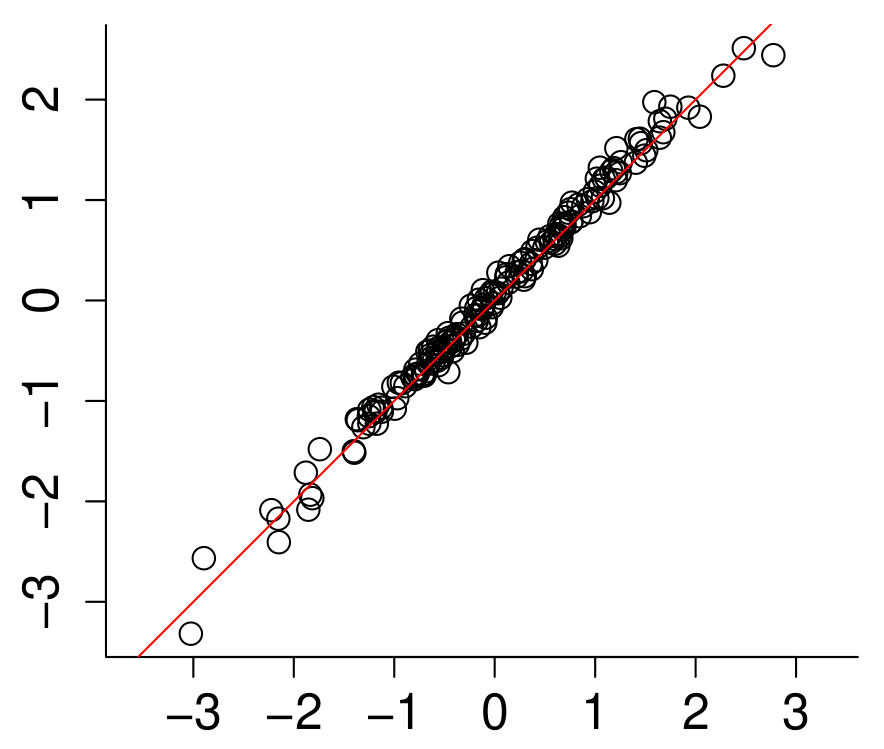}
		\caption{Location - 500.}
	\end{subfigure}
	~
	\begin{subfigure}[b]{0.24\textwidth}
		\includegraphics[width=\textwidth]{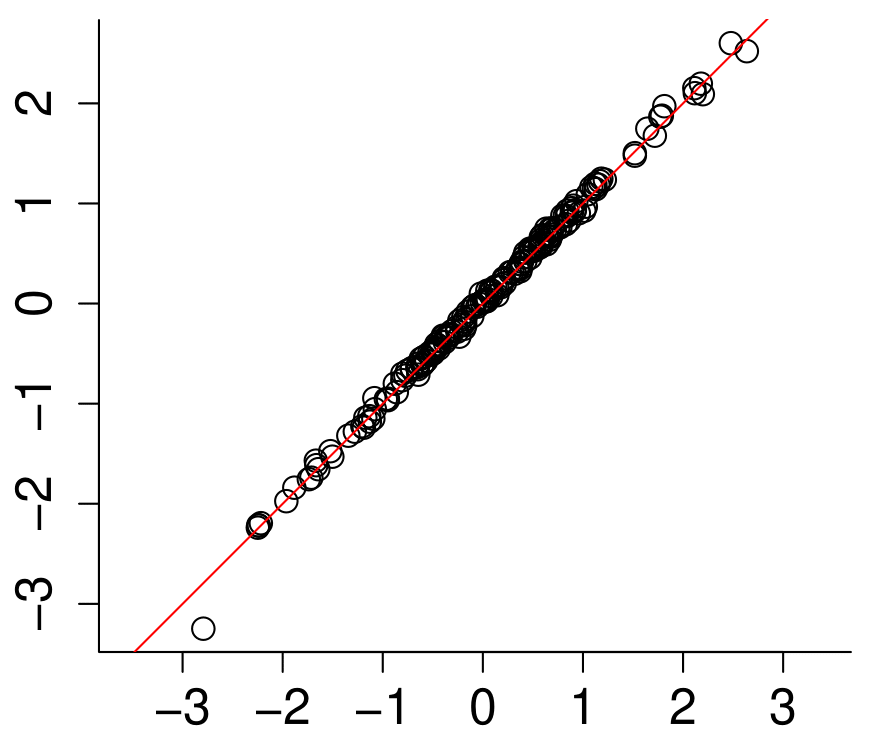}
		\caption{Location - 2k.}
	\end{subfigure}
	~
	\begin{subfigure}[b]{0.25\textwidth}
		\includegraphics[width=\textwidth]{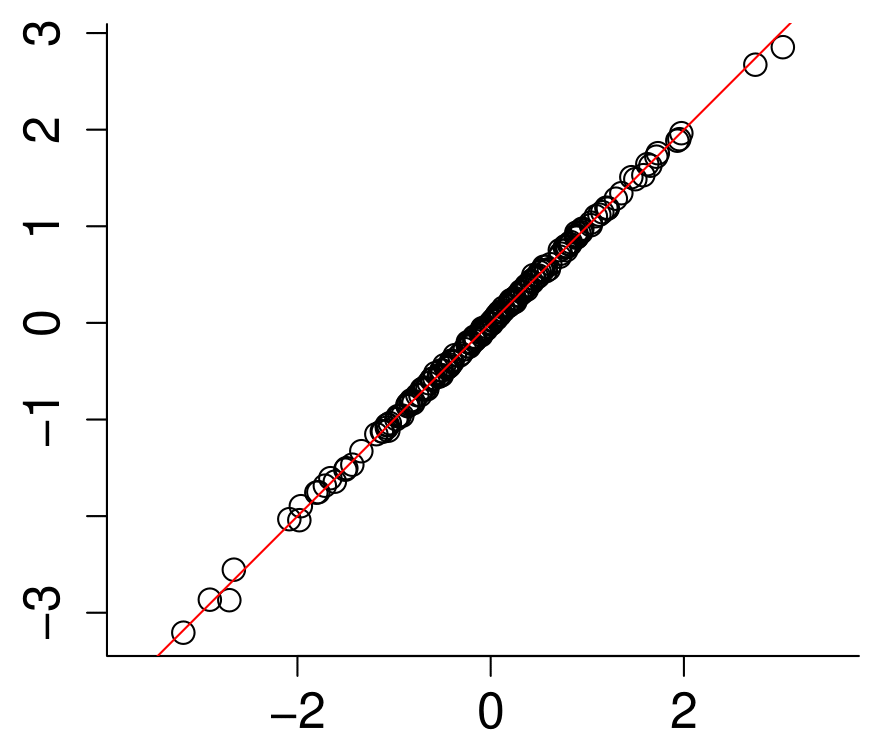}
		\caption{Location - 5k.}
	\end{subfigure}	
	\caption{Posterior mean (y-axis) versus true value (x-axis) (simulations 1-3).}
	\label{graf:Simulacao7_500_ParItens}
\end{figure}

\end{document}